\documentclass[sigconf,10pt]{acmart}

\usepackage[english]{babel}
\usepackage{blindtext}

\renewcommand\footnotetextcopyrightpermission[1]{} 
\setcopyright{none}

\usepackage{multirow}
\usepackage{subcaption}

\usepackage{etoolbox}
\usepackage{hyperref}
\usepackage{xcolor}
\usepackage{caption}
 \usepackage{xspace}
\usepackage{amsfonts}
\usepackage{mathtools}
\usepackage{tabularx}
\usepackage{enumitem}
\usepackage{float}
\usepackage[font=bf,labelfont=small]{caption}
\usepackage[binary-units=true]{siunitx}
\usepackage[ruled, vlined, linesnumbered, commentsnumbered]{algorithm2e} 
\usepackage{pgfplots}
\usepackage{outlines}
\usepackage{comment}

\usepackage{caption}

\setcopyright{none}
\renewcommand\footnotetextcopyrightpermission[1]{} 

\newcounter{packednmbr}
\newenvironment{packedenumerate}{\begin{list}{\thepackednmbr.}{\usecounter{packednmbr}\setlength{\itemsep}{0.5pt}\addtolength{\labelwidth}{-4pt}\setlength{\leftmargin}{\labelwidth}\setlength{\listparindent}{\parindent}\setlength{\parsep}{1pt}\setlength{\topsep}{0pt}}}{\end{list}}
\newenvironment{packeditemize}{\begin{list}{$\bullet$}{\setlength{\itemsep}{0.5pt}\addtolength{\labelwidth}{-4pt}\setlength{\leftmargin}{\labelwidth}\setlength{\listparindent}{\parindent}\setlength{\parsep}{1pt}\setlength{\topsep}{0pt}}}{\end{list}}

\newcommand\blfootnote[1]{%
  \begingroup
  \renewcommand\thefootnote{}\footnote{#1}%
  \addtocounter{footnote}{-1}%
  \endgroup
}

\usepackage{etoolbox}

\newbool{IsPrintComment}
\booltrue{IsPrintComment}   

\newcommand{\name}{Earth+\xspace}

\newcommand{\shadi}[1]
{
	\ifbool{IsPrintComment}
	{
		{\color{magenta}(SN) #1}
	}
	\  
}
\newcommand{\todo}[1]
{
	\ifbool{IsPrintComment}
	{
		{\color{blue} #1}
	}
	\  
}

\newcommand{\kt}[1]
{
	\ifbool{IsPrintComment}
	{
		{\color{olive}(KD) #1}
	}
	\  
}

\definecolor{indigo}{rgb}{0.0, 0.25, 0.42}

\newcommand{\ktedit}[1]
{#1}

\newcommand{\yh}[1]
{
	\ifbool{IsPrintComment}
	{
		{\color{cyan}{\footnotesize [YH: #1]\xspace}}
	}
	\  
}

\newcommand{\peder}[1]
{
	\ifbool{IsPrintComment}
	{
		{\color{blue}(PO) #1}
	}
	\  
}

\newcommand{\junchen}[1]
{
	\ifbool{IsPrintComment}
	{
		{\color{brown}(JJ) #1}
	}
	\  
}

\newcommand{\TODO}[1]
{
	\ifbool{IsPrintComment}
	{
		{\color{red}todo: #1}
	}
	\  
}

\newcommand{\squishlist} 
{
    \begin{list}{$\bullet$}
    {
        \setlength{\itemsep}{0pt}      \setlength{\parsep}{3pt}
        \setlength{\topsep}{3pt}       \setlength{\partopsep}{0pt}
        \setlength{\leftmargin}{1.5em} \setlength{\labelwidth}{1em}
        \setlength{\labelsep}{0.5em}
    }
}

\newcommand{\squishend}
{
    \end{list}
}

\newcommand{\eg}{{\it e.g.,}\xspace}
\newcommand{\ie}{{\it i.e.,}\xspace}

\newcommand{\fillme}{{\bf XXX}\xspace}

\newcommand{\mypara}[1]{\vspace{0.05cm}\noindent{\bf {#1}:}~}

\newcommand{\myparaq}[1]{\smallskip\noindent{\bf {#1}?}~}

\newcommand{\tightcaption}[1]{
\caption{{\normalfont{\textit{{#1}}}}}
}

\newcommand{\tightsection}[1]{
\section{#1}
}

\newcommand{\tightsubsection}[1]{
\subsection{#1}
}

\begin{document}

\title{\name: on-board satellite imagery compression leveraging historical earth observations}


\author{\Large {Kuntai Du$^1$, Yihua Cheng$^1$, Peder Olsen$^2$, Shadi Noghabi$^{2*}$, Ranveer Chandra$^2$, Junchen Jiang$^1$} \\
\textit{$^1$The University of Chicago~~~~~~~~$^2$Microsoft Research}}




\begin{abstract}

With the increasing deployment of earth observation satellite constellations, the downlink (satellite-to-ground) capacity often limits the freshness, quality, and coverage of the imagery data available to applications on the ground.
To overcome the downlink limitation, we present \name, a new satellite imagery compression system that, instead of compressing each image individually, pinpoints and downloads only recent imagery changes with respect to the history reference images.
To minimize the amount of changes, it is critical to make reference images as {\em fresh} as possible.
\name enables each satellite to choose fresh reference images from not only its own history images but also past images of other satellites from an entire {\em satellite constellation}.
To share reference images across satellites, \name utilizes the limited capacity of the existing uplink (ground-to-satellite)
by judiciously selecting and compressing reference images while still allowing accurate change detection. 
In short, Earth+ is the \textit{first} to make reference-based compression efficient, by enabling constellation-wide sharing of fresh reference images across satellites.
Our evaluation shows that \name can reduce the downlink usage by a factor of 3.3 compared to state-of-the-art on-board image compression techniques while not sacrificing image quality, or using more on-board computing or storage resources, or more uplink bandwidth than currently available.

\end{abstract}

\maketitle
\pagestyle{plain}


\blfootnote{$^*$ This work is done while in Microsoft Research.}



\tightsection{Introduction}

Fresh and high-quality satellite imagery is key to many applications,
from digital agriculture~\cite{mulla2013twenty, hank2019spaceborne,planet-digital-agriculture,digital-agriculture1}, environmental monitoring~\cite{manfreda2018use, tucker2000strategies, tralli2005satellite, cunjian2001extracting,environment-monitoring1}, to automatic road detection~\cite{kopsiaftis2015vehicle, van2018spacenet,automatic-road-detection}, and many more.
As a result, large constellations of Low-Earth-Orbit (LEO) earth observation satellites have been deployed~\cite{planet,kodan,serval} 
to capture high-quality imagery for any location multiple times a day~\cite{planet,kodan,serval}.

However, most satellite imagery data captured by these satellites are currently \textit{not} received on the ground due to the limited {\em downlink} (satellite-to-ground) capacity. 
According to a recent estimate, only 2\% of the total image data observed by each satellite can be downloaded to the ground~\cite{kodan}.
Some mission-specific satellites handle the downlink-capacity limitation by filtering images onboard the satellite~\cite{serval,oec} to focus only on mission-specific areas prepaid by the customer.
However, this approach is not sufficient for 
{\em general-purpose} satellite constellations (\eg Sentinel-2~\cite{sentinel-2}, Doves~\cite{planet}), whose goal is to capture and download satellite imagery over wide geographical regions to serve more applications. 

This paper aims to improve {\em onboard compression} for satellite imagery.\footnote{Better imagery compression can also improve filtering-based solutions like~\cite{serval}.}.
We are inspired by the observation that the terrestrial content changes slowly 
between two consecutive satellite visits at the same location~\cite{satroi,virtual-background}.
Thus, to compress a new image, we can compare it with a recent image of the same region, called a {\em reference} image, to detect the geographic {\em tiles} (defined in \S\ref{subsec:up-to-date-reference}) within the region that has changed and then only compress and download the changed tiles. 
Our measurement on Planet dataset~\cite{planet} shows that
without the interference of clouds, only 20\% of the tiles in each image have changed in the previous five days on average,
which ideally can save downlink usage by up-to 5$\times$ (\S\ref{subsec:up-to-date-reference}).

Yet, realizing the {\em reference-based} encoding for onboard imagery compression can be challenging because the reference image should be as \textit{fresh} and contain as little \textit{cloud} as possible~(\S\ref{subsec:up-to-date-reference}).
\ktedit{Typically, the last cloud-free image captured by the same satellite~\cite{planet} can be over 50 days old on average (\S\ref{subsec:up-to-date-reference}).
With such a large time gap, the reference image and the new image may have substantial differences (more than 50\% of the tiles will have significant changes as shown in \S\ref{subsec:up-to-date-reference}), making reference-based encoding less effective.}

We present \name, a \textit{constellation-wide} reference-based encoding system, where the reference images can be selected from historical images of {\em any satellites} in the constellation.
By broadening the set of potential reference images, \name increases the probability of obtaining fresh and cloud-free reference images.
For example, with images from an entire constellation~\cite{planet}, 
cloud-free images can be obtained every 4.21 days on average, instead of every 50 days with one satellite (\S\ref{subsec:constellation-wide}).

\begin{figure}[t!]
    \centering
    \hspace*{-0.5cm} 
    \includegraphics[width=0.53\textwidth]{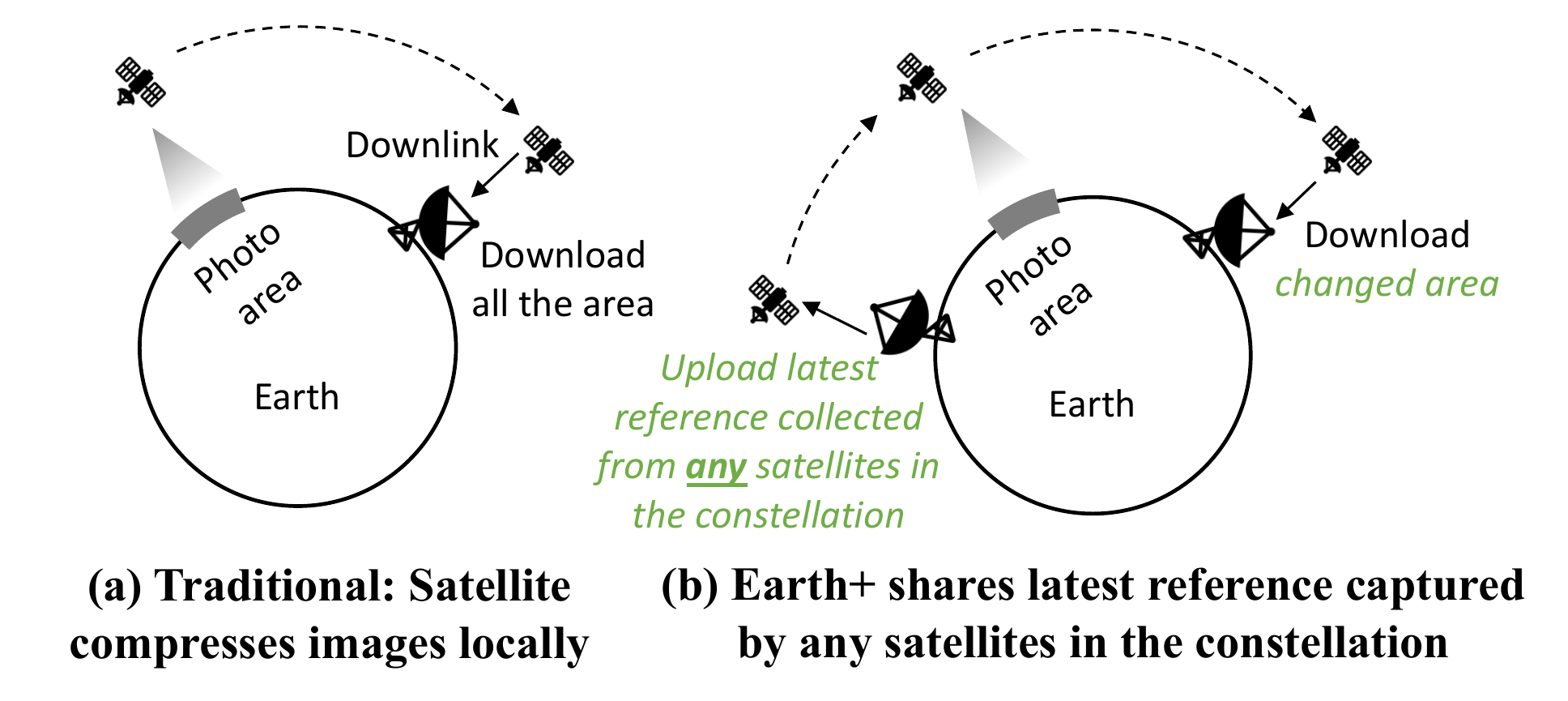}
    
    \tightcaption{Contrasting \name with traditional satellite imagery compression.
    (a) The traditional approach compresses images by satellites using their local onboard information.
    (b) \name's reference-based encoding uses reference images from any satellites in the constellation and uploads the reference images to the satellite to pinpoint the changed areas and only downloads their content.
    }
    \label{fig:highlevel}
\end{figure}

To enable image sharing across satellites in a constellation, \name leverages the existing {\em uplinks} (ground-to-satellite) to \textit{upload} fresh and cloud-free reference images that have been downloaded from different satellites as illustrated in Figure~\ref{fig:highlevel}.
(\S\ref{subsec:workflow} will discuss the rationale of using uplinks, as opposed to alternatives like inter-satellite links.)
The key challenge facing \name is that existing uplinks of earth observation satellites have {\em limited capacity} (\eg 250kbps~\cite{dove-uplink-downlink}). 


We present two techniques (\S\ref{subsec:reference-compression}) to reduce the uplink usage of \name without sacrificing the savings on the downlink.

First, \name uploads reference images at a low resolution while still allowing the satellites to detect the most changed tiles (\S\ref{subsec:reference-compression}).
\ktedit{The rationale is that low-resolution images are sufficient to decide which tiles have changed, which is easier than quantifying how much each pixel in the tile has changed.} 


\ktedit{
Second, \name does not need to store those unchanged tiles when capturing new imagery, which frees up the storage space.
We utilize this freed storage space to cache reference images locally on-board, which allows \name to further reduce the uplink usage by only uploading tiles that have changed relative to the on-board cached reference images.}

Besides the two aforementioned techniques, our implementation of \name (\S\ref{sec:implementation}) also entails techniques to handle satellite-specific issues,
including cloud detection, on-board computation constraints, handling different bands of satellite imagery, and bandwidth variations.

\ktedit{
To put \name's contribution into perspective, while the observation of redundancy across satellite imagery may not be new, prior work on onboard imagery compression has focused on optimizing the compression within each satellite.
In contrast, \name is the first to make reference-based compression efficient, by enabling constellation-wide sharing of fresh reference images across satellites.
}
\ktedit{
We evaluate \name on {\em real-world} satellite specifications (uplink and storage capacities) of the Doves constellation~\cite{dove-constellation} from Planet Labs.
We test \name's compression efficiency on two datasets.
The first dataset is collected from Sentinel-2 dataset~\cite{sentinel-2}, with 3.6 TB data covering 110 thousand km$^2$ from Washington State.
We use this dataset to test \name under a wide range of contents (\eg mountains, forests, and cities), seasons, and under multiple imagery bands (13 bands in total).
Since Sentinel-2 only contains two satellites, we further test \name's performance using the Planet dataset~\cite{planet}, from which we obtain images from 40 satellites for one sampled location (due to the download limit) of 64 km$^2$ in the U.S. for three months. 
Our evaluation shows that:
\begin{packeditemize}
    \item Compared to the state-of-the-art onboard compression schemes, \name reduces the downlink bandwidth usage by 1.3-3.3$\times$ without hurting the imagery quality on all bands.
    This can reduce the reaction delays of ground applications (\eg forest-fire alerts) by upto 3$\times$.
    \item These improvements are achieved {\em without} using more uplink bandwidth than currently available or more compute or storage resources than the baselines.
    \item With more satellites in a constellation, \name can further reduce the amount of downlink bandwidth usage.
\end{packeditemize}

}

\ktedit{
That said, \name's reference-based encoding is not a good fit for applications that require lossless satellite imagery (\S\ref{sec:limitation}). 
}

\textbf{This work does not raise any ethical issues.}



\tightsection{Motivation}
\label{sec:characterization}

We start with the background on satellite imagery and earth observation satellite constellations.

\tightsubsection{Background}
\label{subsec:satellite-connectivity}

Many applications can benefit from frequently updated (\eg daily) and high-resolution satellite imagery.
For example, precision agriculture ideally needs daily access to satellite imagery with each pixel corresponding to a $5\mathrm{m}\times5\mathrm{m}$ area on Earth~\cite{agriculture-daily,agriculture-daily2} to help timely decisions on the distribution of fertilizers, pesticides, and water. 
Also, wildfire monitoring requires the imagery to be updated frequently with sufficient resolution to promptly detect and respond to fire outbreaks, mitigating potential damage~\cite{serval}.

To provide fresh, high-resolution satellite imagery, many LEO satellites (\eg >100 satellites~\cite{planet}) are deployed to form satellite constellations.
We characterize two features:
\begin{packeditemize}
    \item \textit{High-resolution imagery}: LEO satellites are close to the ground (due to their low earth orbits) and can capture imagery with low ground-sampling distance (GSD for short, lower GSD means higher resolution).
    \item \textit{Frequent revisit}:
    With a large number of satellites, any location on the earth's surface will be frequently revisited (\eg daily ~\cite{planet}), while a single satellite can only revisit one location once every ten days~\cite{sentinel-2-revisit}.
\end{packeditemize}

\begin{figure}
    \centering
    \includegraphics[width=0.5\columnwidth]{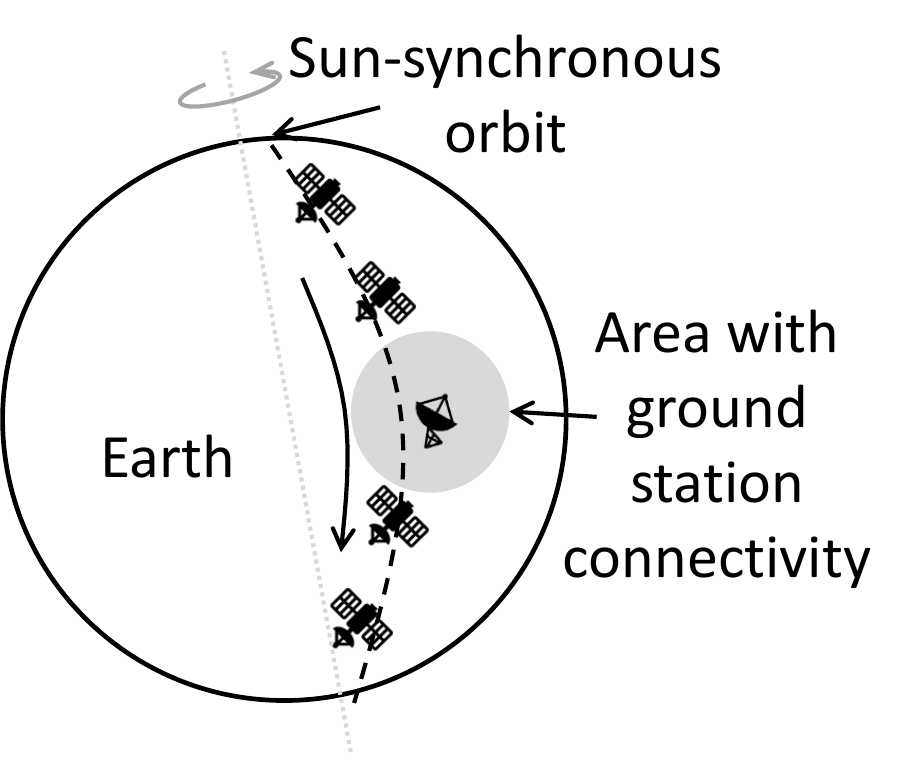}
    \tightcaption{Illustration of a LEO satellite constellation: the satellites follow a sun-synchronous orbit, and the ground station downloads imagery data from a satellite when the satellite passes through it.
    }
    \label{fig:leo-orbit-illustration}
\end{figure}

Figure~\ref{fig:leo-orbit-illustration} shows an illustrative example of such a large-scale LEO satellite constellation, where multiple satellites are located in a sun-synchronous orbit\footnote{A sun-synchronous orbit ensures that each location on the earth is approximately revisited once per day, at approximately the same time local time, allowing the constellation to capture images with similar illumination condition once per day.} and these satellites can potentially stream data to the ground when they are close enough to one of the ground stations (we only plot one ground station in the figure for simplicity).

Note that in the following text, we denote \textit{\textbf{the ground}} as the ground stations that the constellation can potentially contact and the computation and networking infrastructures around these ground stations.

\tightsubsection{Downlink bottleneck and our objective}

\mypara{Downlink capacity gap}
Despite more images being captured by the satellites, only a small fraction of data are downloaded to the ground due to 
the limited capacity of the downlink (satellite-to-ground).\footnote{To put this gap in perspective, we use the data from the Doves constellation in 2017~\cite{planet}, where each satellite image covers an area of 400~km$^2$(assuming 6600×4400 resolution~\cite{planet-ground-sampling-distance} and a GSD of 3.7 meters~\cite{planet-ground-sampling-distance}) so after scanning the whole earth's surface, we would accumulate 1.275 million images or 191.25 TB of data (with each image being 150~MB~\cite{planet-image-size}).
To estimate how much data can be downloaded, we make the idealized assumption that the satellite always has a stable connection of 200~Mbps~\cite{dove-uplink-downlink} with a ground station. 
Now, each satellite scans the whole Earth's surface every ten days~\cite{sentinel-2-revisit}, so under the idealized assumption, the satellite can only transmit 21.6~TB of data to the ground, which is less than 12\% of the 191.25~TB of data accumulated during that period of time.}
Specifically, we refer to {\em downlink bandwidth} as the average download speed from satellites to the ground during each ground contact.
The exact gap between the downlink capacity and the imagery data varies with the constellation, and a recent study shows only about 2\% of the images captured by satellites are actually downloaded to the ground~\cite{kodan}.

Further, the downlink demand is constantly growing, with higher resolution (\eg a GSD of 0.5m~\cite{high-resolution-satellite-imagery}) and more bands found to be useful (\eg vegetarian red edge band and water vapor band~\cite{sentinel-2-bands}).
In contrast, the downlink grows slowly due to the long deployment cycle of satellites.
These trends suggest that the gap between the demand for downlink bandwidth and its actual capacity will likely persist if not increase.

\mypara{Optimization objective}
We aim to address the downlink bottleneck of satellite constellations by better {\em satellite image compression}.
More specifically, we aim to use much less bandwidth to download the same amount of satellite imagery, measured in the number of photoed locations and frequencies, without compromising image quality.
To measure the quality of the downloaded images, we use Peak Signal-to-Noise Ratio (PSNR for short), which aligns with satellite imagery compression literature~\cite{indradjad2019comparison,gunasheela2018satellite,faria2012performance,shihab2017enhancement}.

\mypara{On-board constraints}
While optimizing for the image quality and reducing the downlink consumption, we stick to real-world on-board storage, computation, and uplink constraints. 
We describe the real-world satellite specification that we used for our evaluation in \S\ref{sec:evaluation}.

\tightsubsection{Existing solutions}

There are several approaches to addressing the downlink bandwidth bottleneck.

The first is to physically increase the downlink capacity by upgrading the infrastructure, \eg building more ground stations~\cite{planet} or adding more satellites to parallelize the downloading of multiple satellites~\cite{planet,sentinel-2,kodan}.
The costs of such infrastructure changes can be prohibitive, and they can be slow. For context, it takes tens of millions of dollars to build and send just one single satellite~\cite{satellite-cost}.

An alternative is to filter the imagery onboard the satellite~\cite{kodan,serval,oec}.
For the mission-specific constellations that focus on specific regions, this approach can filter out most of the imagery.
For instance, the Biomass mission targets forest areas to monitor forest coverage changes~\cite{biomass}, while the IceBridge mission observes polar ice to gauge climate change impacts~\cite{icebridge}. 
However, 
they must exclude data useful for other applications. For example, the Biomass mission omits about 91\% of the Earth's surface~\cite{forest-area,ocean-area}, such as city areas (which are useful for smart city applications) and agriculture areas (useful for digital agriculture).

This work focuses on the third approach: onboard imagery compression. 
It is complementary to the first two approaches. 
Existing solutions include augmenting single-image codecs~\cite{edge-prediction,progressive-jpeg2000,CALIC,linear-prediction,band-reordering,band-reordering-segmentation,band-reordering-prediction,band-prediction} and developing more expensive neural-based codecs such as autoencoders~\cite{autoencoder-1,autoencoder-2,autoencoder-3,autoencoder-4,autoencoder-5}.

However, most techniques focus on compressing single imagery from a single satellite, so they fall short in leveraging the {\em redundancies between} images to achieve higher compression efficiency.

\section{Reference-based encoding}
\label{subsec:up-to-date-reference}



Next, we introduce reference-based encoding, a seemingly promising idea that leverages a reference image to pinpoint and download only regions that have recently changed. 
As we will see, directly applying this approach to a satellite does not work well as images locally available to each satellite may not be recent enough or contain too much cloud to realize the benefit of reference-based encoding.


\mypara{Background on reference-based encoding}
Reference-based encoding is commonly used to compress a sequence of images whose content changes slowly and gradually with respect to time\cite{h264,h265,vp8,vp9,satroi,virtual-background}, such as video streams.
Existing reference-based encoding systems (\eg video codecs~\cite{h264,h265,vp8,vp9}) typically select some of the images as the {\em reference} and encode the remaining images by encoding their difference concerning the reference images.
As existing codecs encode the images at the granularity of \textit{tiles} (a tile is a block of pixels, where we use a 64$\times$64 pixel block as a tile by default), and the difference is separately calculated per tile. 

Since the satellite imagery captured for the same location also changes slowly over time (as shown in prior work~\cite{satroi,virtual-background}), there is some recent work to apply reference-based encoding in onboard satellite imagery compression~\cite{satroi,virtual-background}.
Given a new image, it compares the image with a reference image of the location from the past and pinpoints the changed tiles with a pixel difference greater than the threshold compared to the reference. 
It then encodes those changed tiles and downloads the tiles in their entirety.\footnote{Unlike conventional video frame encoding, the changed tiles are downloaded in their original pixel values rather than the pixel differences between the new image and the reference. This is mainly because the reference images are downsampled due to limited onboard storage, thus encoding the tile itself and the difference concerning its reference requiring a similar amount of bits.}
Our work follows this approach when encoding changed tiles (\S\ref{sec:implementation}).

\begin{figure}
    \centering
    \begin{subfigure}[b]{0.32\columnwidth}
         \centering
         \includegraphics[width=\textwidth]{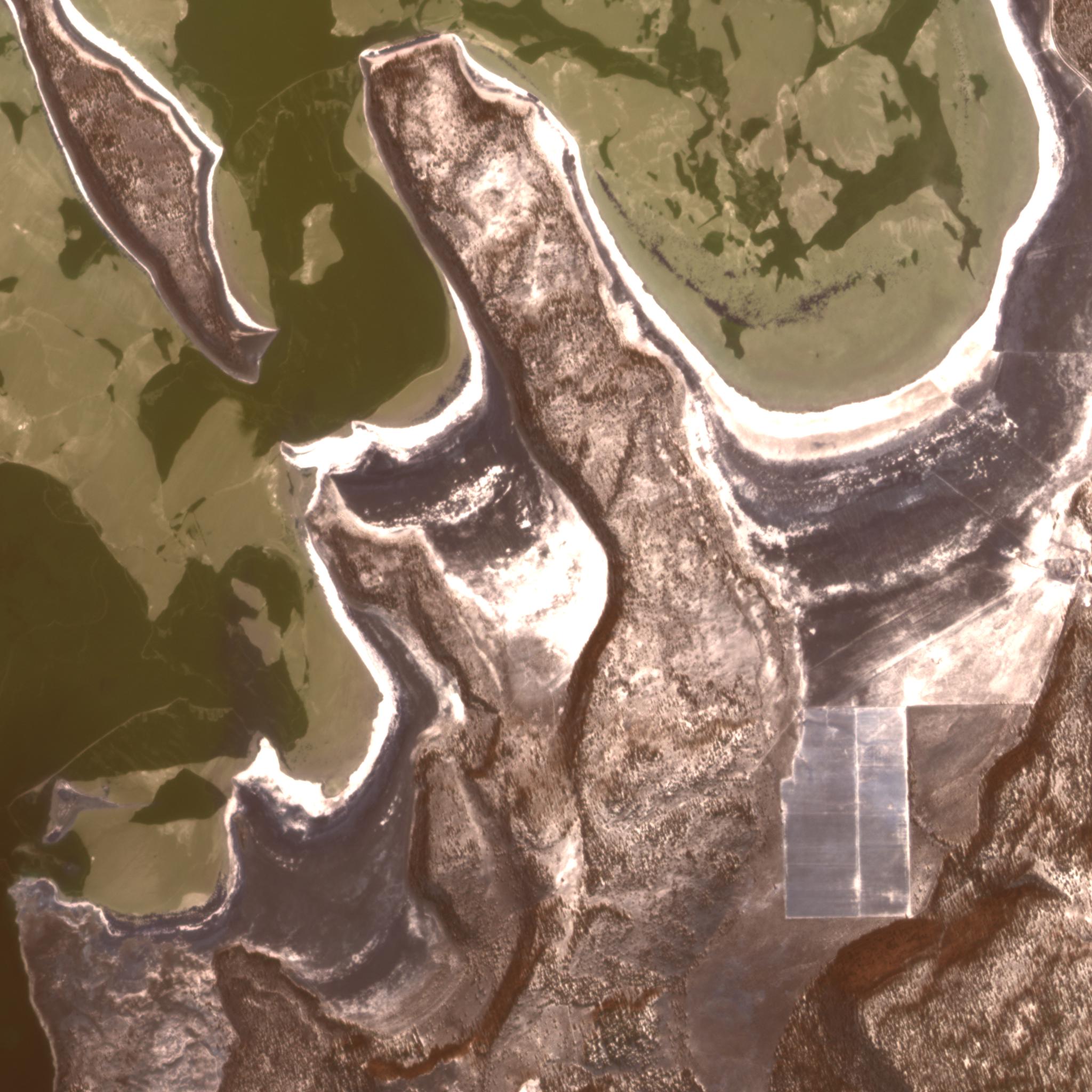}
         \caption{Captured image (Day 30)}
     \end{subfigure}
    \hfill
    \begin{subfigure}[b]{0.32\columnwidth}
         \centering
         \includegraphics[width=\textwidth]{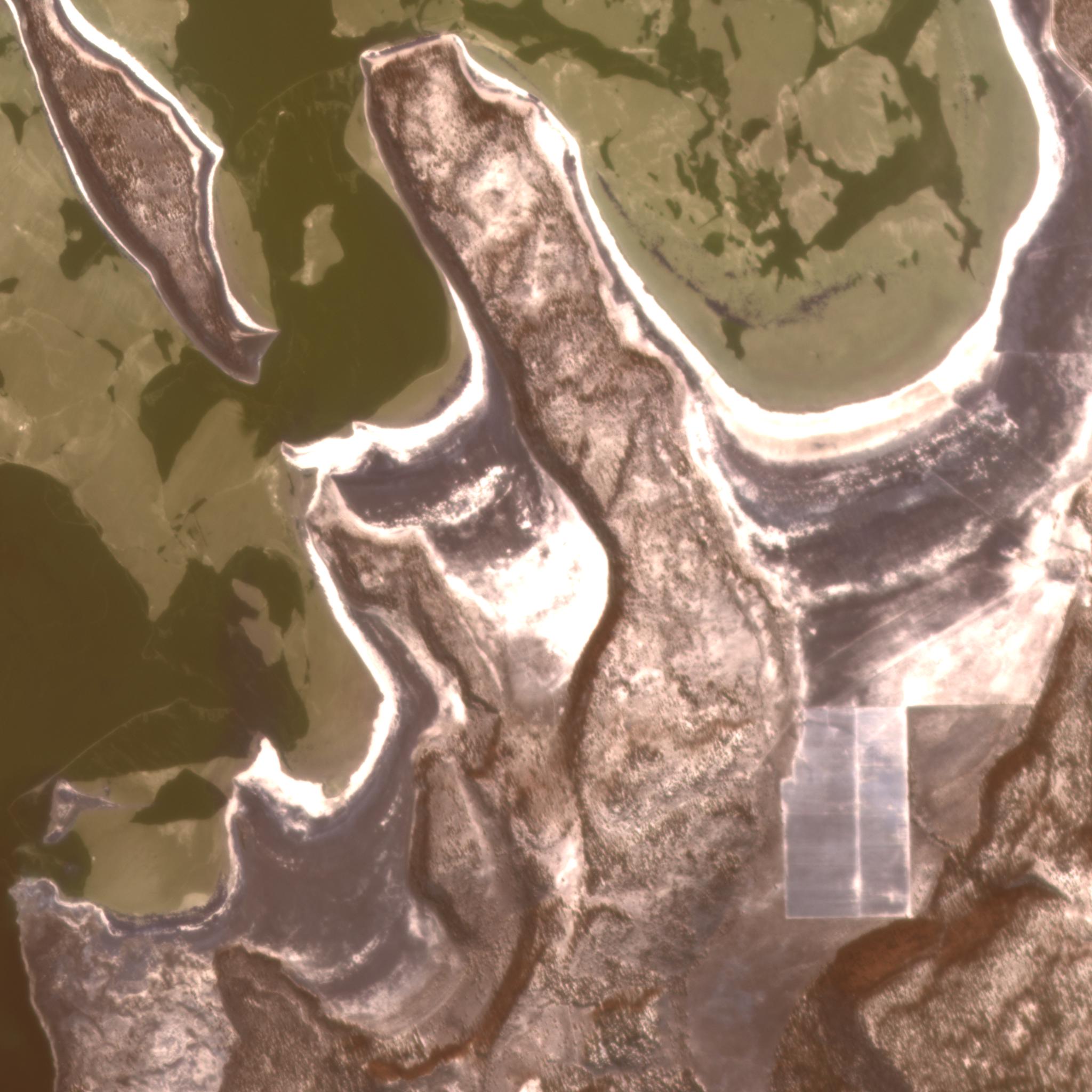}
         \caption{Fresh reference (Day 27)}
     \end{subfigure}
    \hfill
    \begin{subfigure}[b]{0.32\columnwidth}
         \centering
         \includegraphics[width=\textwidth]{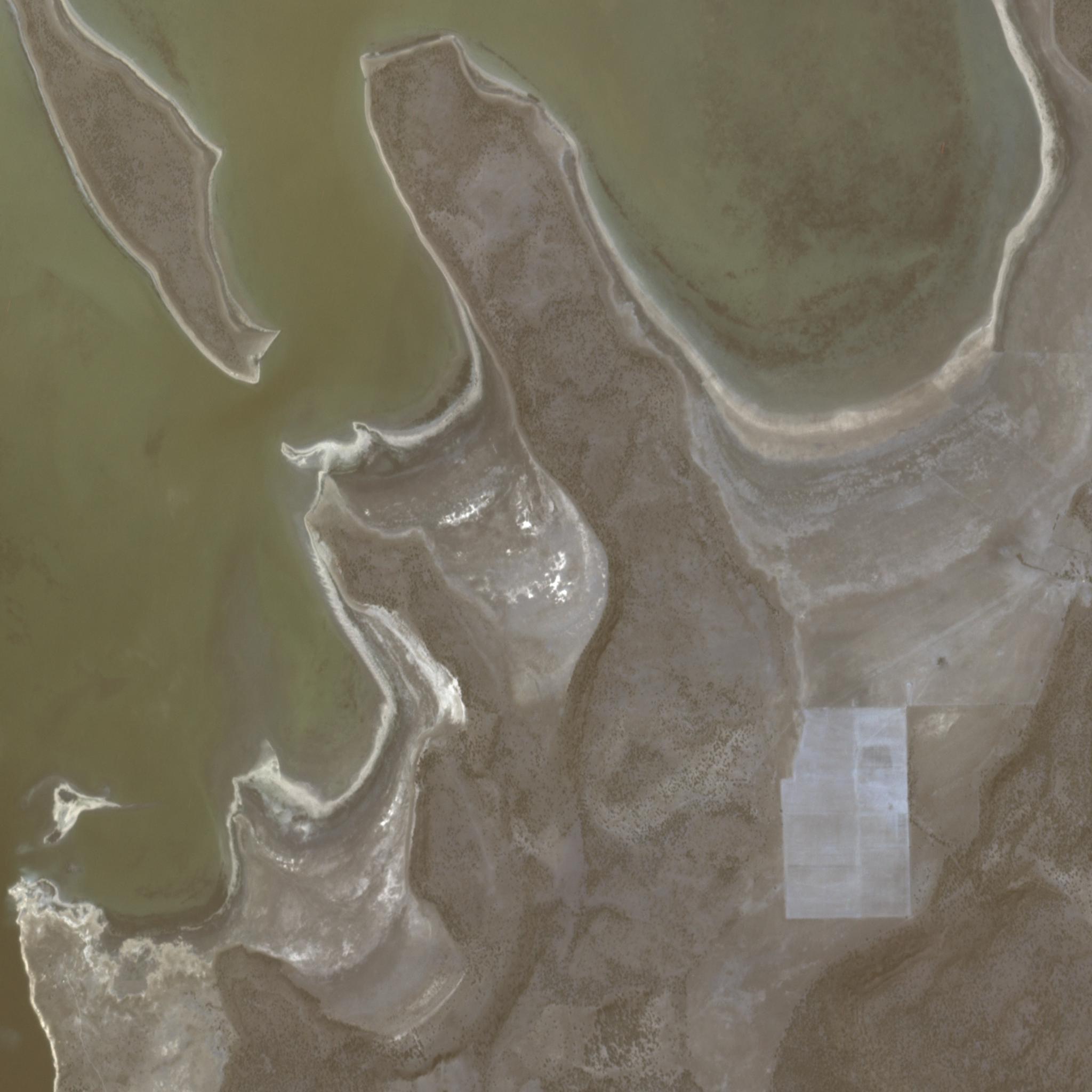}
         \caption{Older reference (Day 1)}
     \end{subfigure}

    \tightcaption{An example illustrating why reference images need to be fresh. Comparing a captured image from Day 30 with a fresh reference from Day 27 reveals much fewer changes than when comparing it with an older reference from Day 1. Image © 2023 Planet Labs PBC.
    }
    \label{fig:illustrative-age}
\end{figure}

\begin{figure}[t!]
    \includegraphics*[width=0.75\columnwidth]{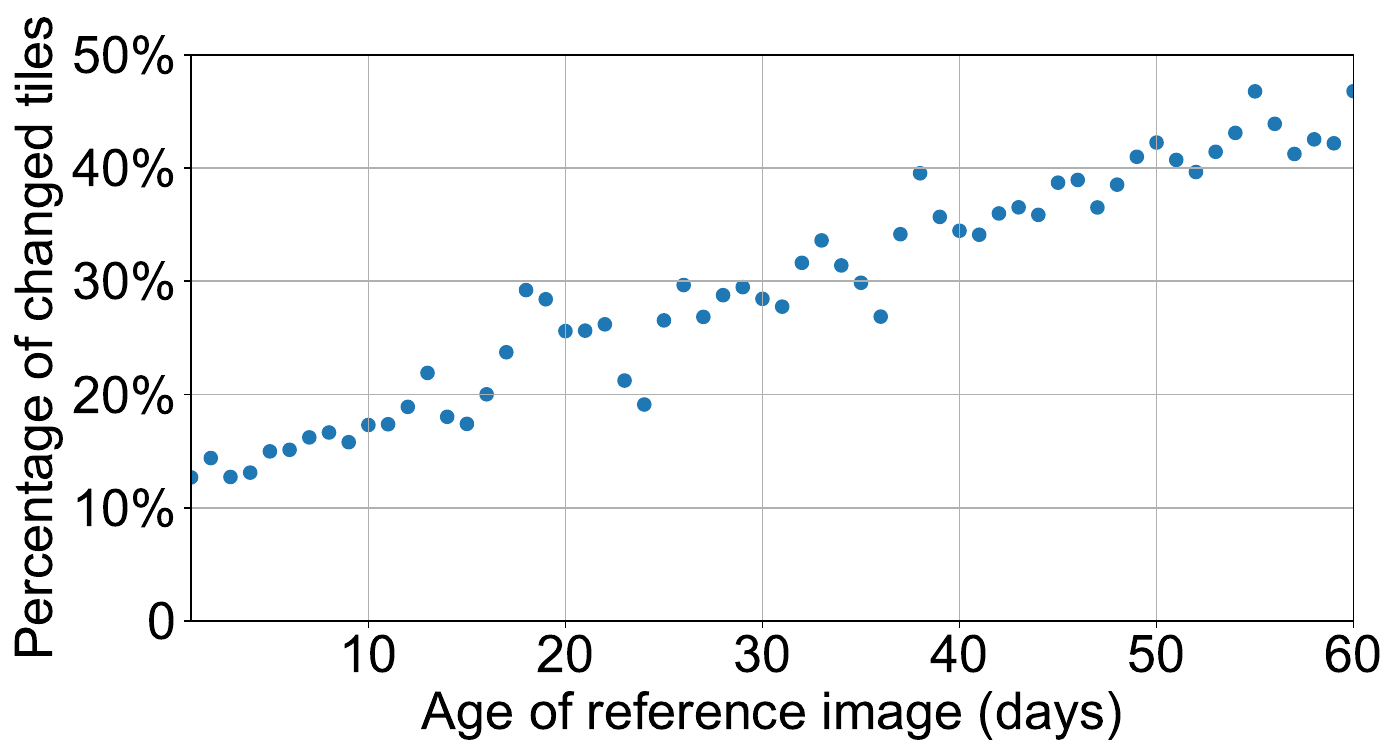}
    \tightcaption{More changes need to be downloaded when the age of the reference image gets larger.}
    \label{fig:staleness-change}
\end{figure}

\mypara{Reference images need to be {\em fresh}}
While reference-based encoding seems to be a good fit for imagery compression, it is only effective if the {\em age} of the reference image---the time gap between the reference image and the currently observed image---is as low as possible. 
Reference image with high age leads to more changed areas in the currently observed image, which must be downloaded to the ground. 
Figure~\ref{fig:illustrative-age} provides an illustrative example, where the changes that happened on Day 3 compared to Day 1 are much less than on Day 30 compared to Day 1.
To make it more concrete, we use three months of cloud-free (explained shortly) images from the Planet dataset~\cite{sentinel-2} on one randomly sampled location in the U.S.
Here, we say a tile has changed if it has an average pixel differences greater than 0.01 after aligning the illumination (\S\ref{sec:implementation}).\footnote{The pixel differences are computed after we normalize pixel values to [0,1]. The threshold of 0.01 means that those areas deemed ``unchanged'' would be above 40, which is very high in satellite imagery compression literature~\cite{psnr-38-high} and provides almost the same results as the uncompressed image in applications like satellite imagery compression~\cite{psnr-38-classification}. }
Figure~\ref{fig:staleness-change} shows a steady increase in the percentage of changed areas with the age of the reference image: \textit{the percentage of changed tiles will increase by 3$\times$ if increasing the age of the reference image from 10 days to 50 days}.


\mypara{Reference images should be {\em cloud-free}}
If some tiles in the reference image are covered by clouds, they are not useful as a reference to detect changes. As a consequence, the corresponding tiles in the current image can only be deemed as changed and downloaded to the ground.
This greatly compromises the benefit of reference-based encoding.

\begin{figure}
    \centering
    \includegraphics[width=0.75\columnwidth]{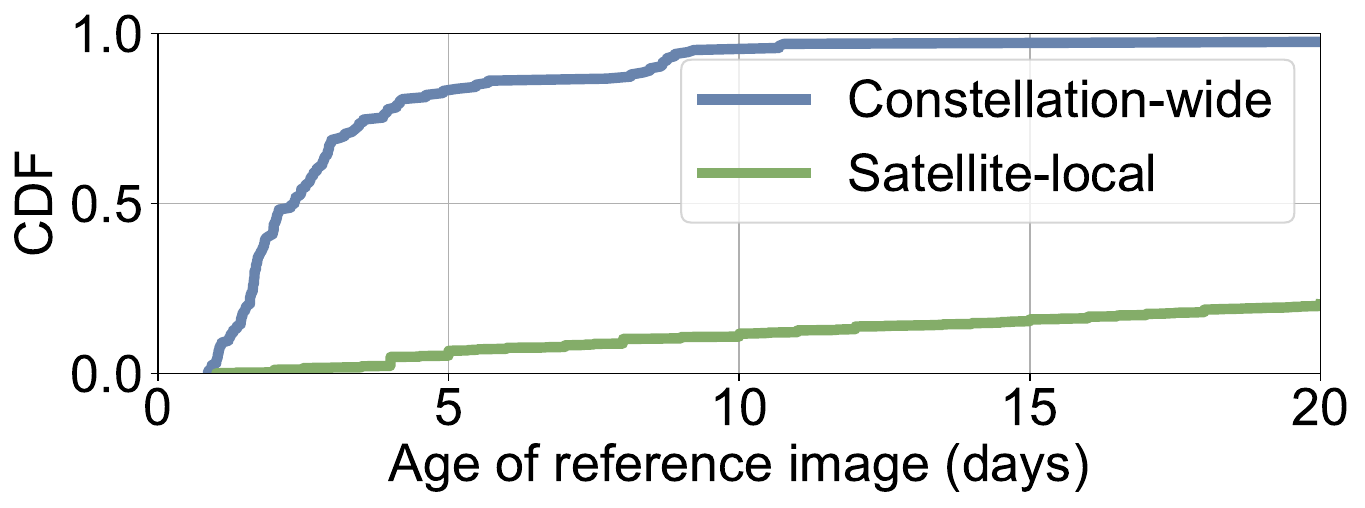}
    \vspace{-0.3cm}
    \tightcaption{Measuring the age of reference images under two strategies: updating the reference using historical images locally captured by the satellite (``Satellite-local'') and updating the reference using images from any satellites in the constellation (``Constellation-wide''). 
    It shows that the constellation-wide approach can reduce the average age of the reference image \textbf{from 51 days to 4.2 days}, a \textbf{12}$\times$ reduction. 
    }
    \label{fig:age-measurement}
\end{figure}

\myparaq{Why reference-based encoding is challenging}
In practice, however, there may not always exist a reference in the satellite's history images that is {\em both} fresh and covered by little cloud.
For example,
existing work~\cite{satroi,virtual-background} stores a fixed reference image on-board, which will get older over time and make most of the areas being counted as changed and downloaded to the ground, negating the benefit of reference-based encoding.
Moreover, \textit{even if} a satellite \textit{were} able to choose the reference image from all of its historical images, the most recent reference image with less than 1\% cloud coverage would still be tens of days old.
For instance, Figure~\ref{fig:age-measurement} shows the age distribution of the closest reference images that are covered by less than 1\% cloud if the satellite chooses the reference image by itself (\ie the ``Satellite-local'' curve in the figure). 
We note that the age of the most recent cloud-free reference image is 51 days on average. 
The reason for the high ages of recent cloud-free images is two-fold:
\begin{packeditemize}
    \item A single satellite revisits the same location at a low frequency (once every 10-15 days~\cite{sentinel-2-revisit}). This is because LEO satellites can only capture a small area on Earth at a time (since their size is small~\cite{serval} and they are close to Earth), necessitating extended periods to complete a full scan of the Earth before revisiting the same locations.
    \item Since, on average, 2/3 of the earth is covered by clouds~\cite{cloud-coverage}, so even if the most recent image of the same location is ten days old, it may likely be (partly) covered by cloud and are not ideal choice for reference images.\footnote{Compared to picking a complete image as the reference, choosing a different reference for each tile may lower the age of the reference, but in practice this reduction is marginal because when the cloud is present, it often covers most of an image and the remaining content, if any, will be influenced by its shadow, making change detection difficult.}
\end{packeditemize}

\begin{figure*}[ht!]
     \centering
     \includegraphics[width=1\textwidth]{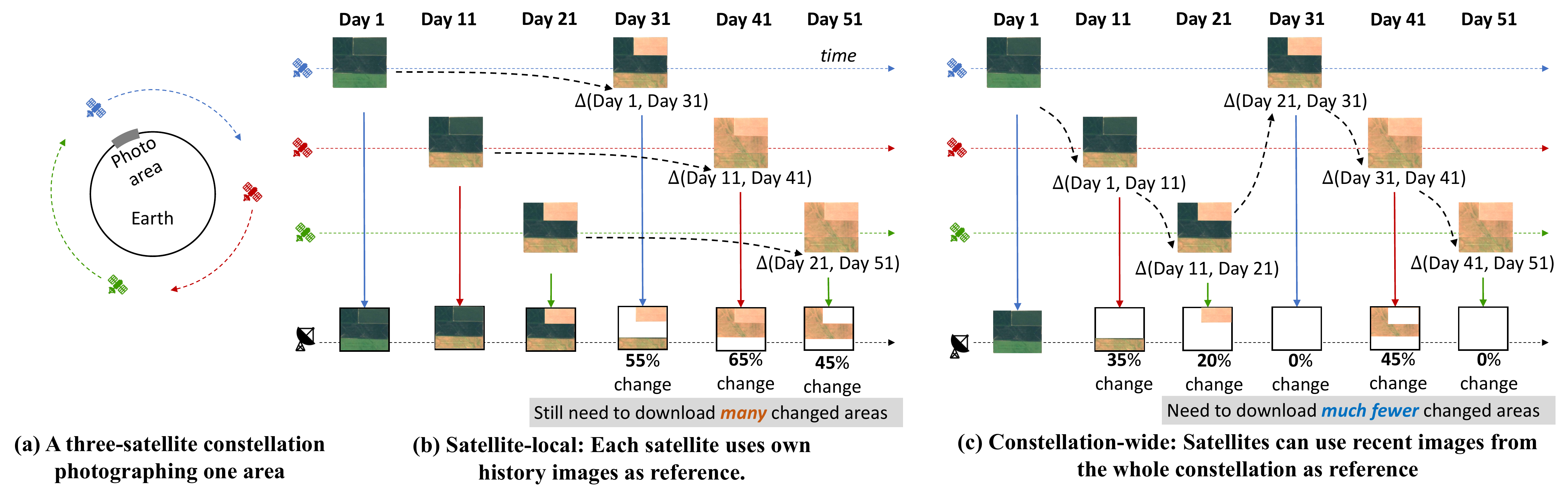}
     \vspace{-0.3cm}
     \tightcaption{Contrasting two reference image uploading strategies: (a) using satellite-local images as reference images and (b) uploading reference images using uplink. By uploading images using uplink, one can reduce the average age of the reference image by 3$\times$ (from 30 days to 10 days) and the downloaded areas by 3.6$\times$ (55\% to 15\%) in this example.}
     \label{fig:uplink-design}
\end{figure*}

\section{\name: Constellation-wide Reference-based encoding}

To improve onboard satellite imagery compression, we present \name, a reference-based encoding system that obtains fresh and cloud-free reference images from images captured by any satellites in the \textit{\textbf{whole constellation}}, rather than the history images of the same satellite.
This section introduces the idea of constellation-wide reference sharing (\S\ref{subsec:constellation-wide}) and an overview of \name (\S\ref{subsec:workflow}).
We then present the design of \name that makes constellation-wide reference-based encoding practical (\S\ref{subsec:reference-compression}).

\tightsubsection{\textit{Constellation-wide} reference selection}
\label{subsec:constellation-wide}

Compared to the prior work, which only refers to local images observed by the same satellite, \name \textit{augments} the set of reference images that reference-based encoding can choose from and thus potentially reduces the age of reference images, leading to fewer changes to be downloaded to the ground. 

To illustrate the benefits and challenges of \name, we contrast two designs.
\begin{packeditemize}
    \item {\em Satellite-local reference:} Pick the latest cloud-free image observed by the same satellite as the reference image.
    \item {\em Constellation-wide reference:} Pick the latest cloud-free image observed by \textbf{\textit{any satellite in the whole constellation}} as the reference image.
\end{packeditemize} 
Note that the latter is not practical because it needs a large amount of bandwidth to share the reference images, a challenge we will tackle soon in \S\ref{subsec:reference-compression}.

Figure~\ref{fig:uplink-design} gives an illustrative example of this contrast with a constellation of three satellites (in different colors).
The goal is to compress images taken by these satellites for the same location.
To simplify the discussion, all images in this example are cloud-free.
Each satellite takes a cloud-free image every 30 days, so the satellite-local reference (Figure~\ref{fig:uplink-design}(b)) will be 30 days old.
Consequently, in the last three images (Day 31, 41, and 51), 45\%-65\% of tiles are deemed as changed and need to be downloaded. 

In contrast, with constellation-wide reference Figure~\ref{fig:uplink-design}(c)), since the reference image can be from any satellite, the freshest reference is only ten days old rather than 30 days. 
As a result, two of the three last images do not have any changed tiles and one has only 45\% changed tiles, \ie only 15\% are changed tiles on average.
In short, the ability to pick reference images from any satellite in the constellation reduces the age of reference images by 3$\times$ (30 days to 10 days) compared to the satellite-local design, and this reduces the changed tiles to download by 3.6$\times$ (55\% area to 15\% area). 

\tightsubsection{\name workflow}
\label{subsec:workflow}

\name is a concrete design of constellation-wide reference-based encoding.
It answers two basic questions: (1) which reference images should be shared between different satellites, and (2) how to share these reference images using the existing infrastructure.

\begin{figure}[t!]
    \centering
    \hspace*{-0.5cm} 
    \includegraphics[width=0.48\textwidth]{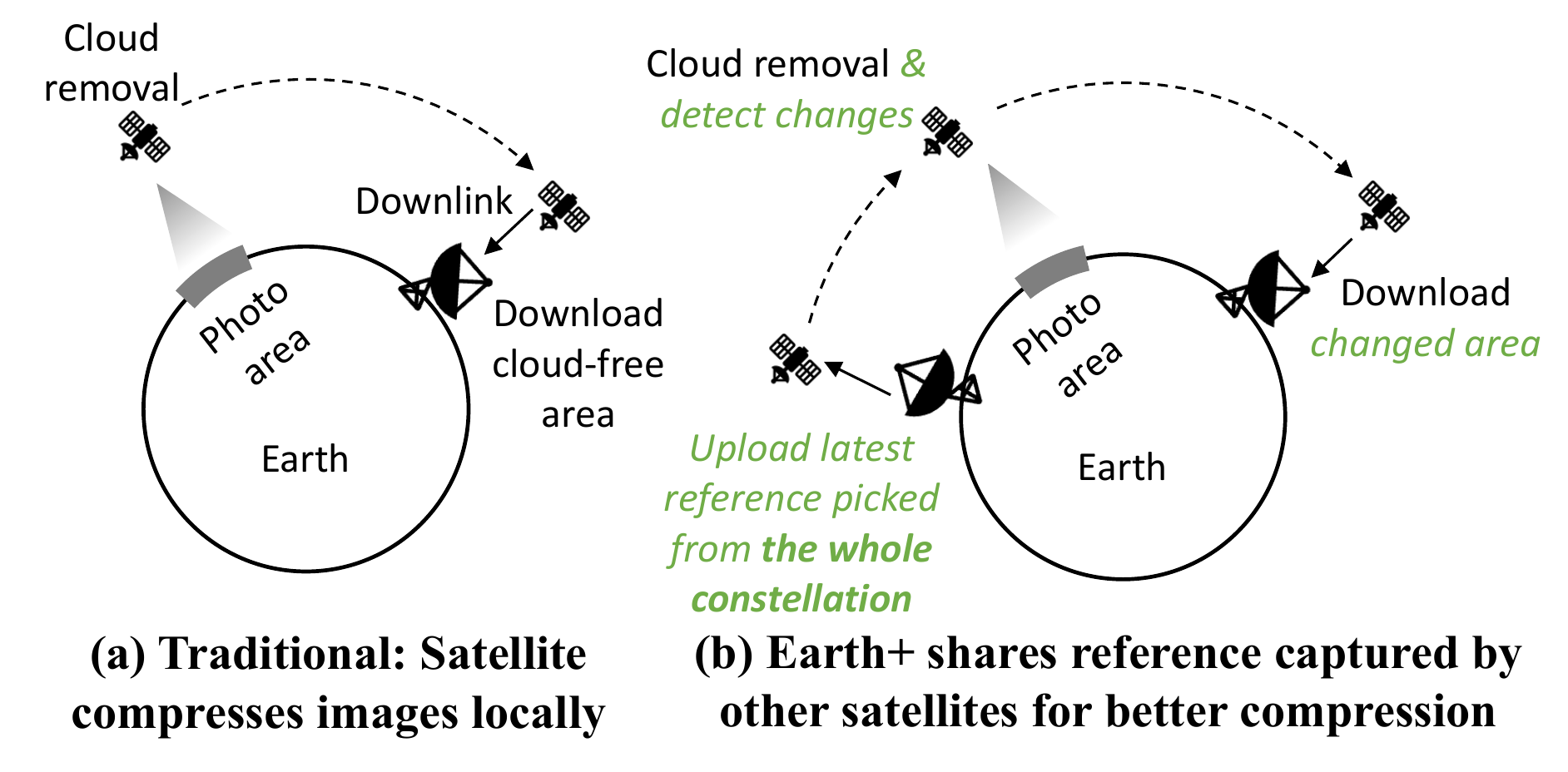}
    \vspace{-0.3cm}
    
    \vspace{-0.3cm}
    \tightcaption{Contrasting between the satellite imagery compression workflow of prior work~\cite{kodan} and \name.
    }
    \label{fig:workflow}
\end{figure}

To answer the first question, \name reuses the images downloaded to the ground from all satellites and {\em selectively uploads} these images as reference images to the satellites.
Figure~\ref{fig:workflow}(b) illustrates this workflow.
\begin{packeditemize}
\item During \textit{\textbf{previous}} ground contact, the ground station \textbf{\textit{uploads}} latest cloud-free images (that can come from \textit{any} satellite in the constellation) as reference images for the locations that the satellite will fly by before the next ground contact\footnote{
It is feasible to upload reference images for geographical locations that the satellite will visit in the future, as these locations can be accurately predicted by, for example, Two Line Element data available in Celestrak~\cite{celestrak}.}.
\item When passing over a location, the satellite captures the imagery, removes clouds, detects changes using the reference images, and encodes the changes.
\item During the next ground contact, the satellite downloads the encoded changes to the ground.
\end{packeditemize}




Compared to the workflow of traditional satellite imagery processing pipelines, which capture images and download them to the ground (as depicted in Figure~\ref{fig:workflow}(a)), \name \textbf{\textit{uploads}} the reference images from the ground to the satellite.
We rely on ground stations as an ``overlay'' point to share images downloaded from each satellite with other satellites.
The rationale is two-fold:
\begin{packedenumerate}
    \item The ground stations can access any historical image observed by the whole constellation, allowing \name to select reference images constellation-wide.
    \item The ground station has sufficient computing resources to more accurately detect clouds and upload only cloud-free images to satellites as the reference (\S\ref{subsec:up-to-date-reference}).
\end{packedenumerate}

A seemingly promising alternative to enable constellation-wide reference is to let satellites share data via inter-satellite links (ISL).
\name does not use ISL because it is currently not available for earth observation satellites~\cite{serval}.
Further, the scale of existing earth observation constellations (less than 200 satellites) is insufficient to guarantee a stable ISL connection between any two satellites, as providing such a guarantee typically requires thousands of satellites (\eg Starlink~\cite{starlink}).

\subsection{Tackling limited uplink bandwidth}
\label{subsec:reference-compression}

However, using the uplink to upload reference images to the satellites is not without challenges---the uplink has limited bandwidth (\eg only 250~Kbps in DOVEs constellation~\cite{dove-uplink-downlink}). 
\name tackles this challenge with three ideas.
Put together, they allow enough reference images to be sent to the satellites under the limited uplink bandwidth while allowing \name to realize sizable downlink savings.

\mypara{Downsampling reference images}
First, \name compresses reference images by downsampling (\ie lowering resolution).
Consequently, \name also detects changed tiles at a low resolution.
For instance, if the original resolution of the captured image is 4000$\times$4000, and the uploaded reference image is downsampled to 500$\times$500, then the satellite will also downsample the captured image to 500$\times$500 before calculating the pixel difference in each tile. 
(Here, we assume both images have been preprocessed by cloud removal and illumination alignment, which will be described in \S\ref{sec:implementation}.)
We then mark the tiles with average pixel difference over a threshold $\theta$ (explained shortly) as changed tiles.
Finally, the changed tiles will be saved and compressed before being sent to the ground. 
Please refer to \S\ref{sec:implementation} on how \name encodes these tiles.


\begin{figure}
    \centering
    \includegraphics[width=0.8\columnwidth]{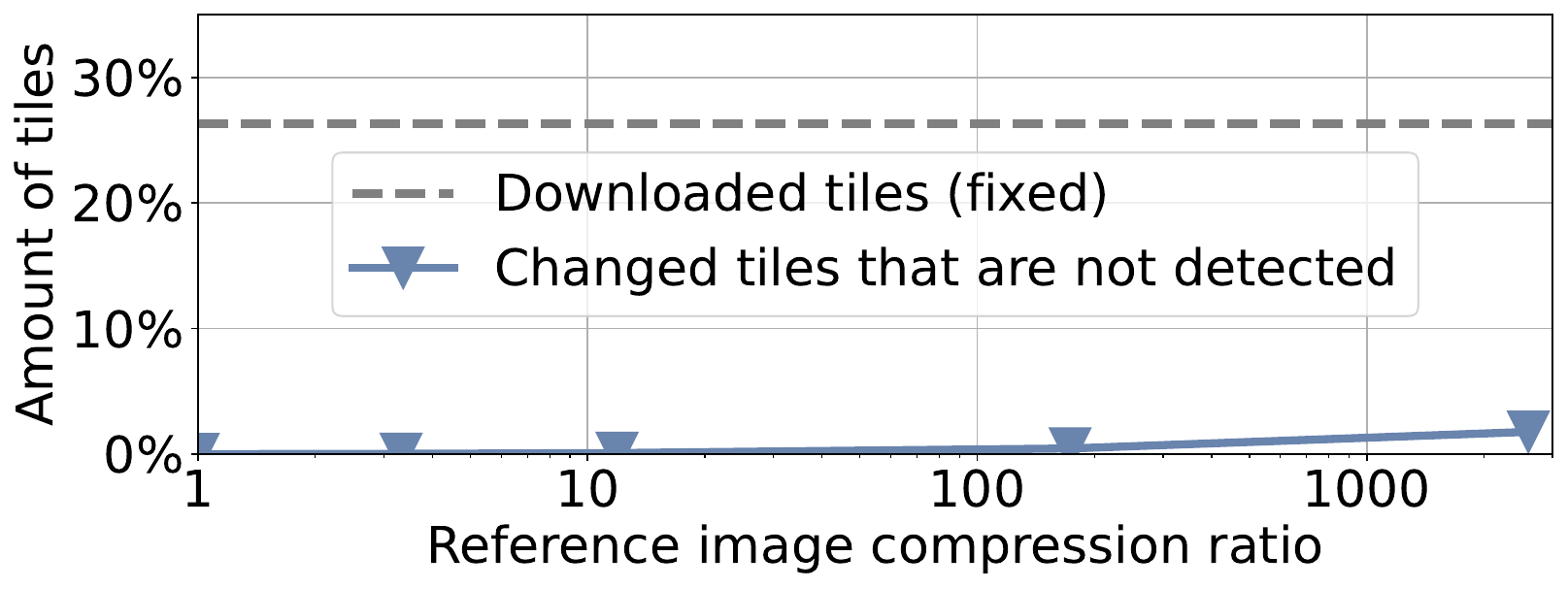}
    \vspace{-0.3cm}
    \tightcaption{
    An example measurement based on the Planet dataset showing how many changed tiles that are undetected given different reference image compression ratios, while fixing the total number of tiles being downloaded.
    It highlights that only 1.7\% of the tiles failed to be detected as changed when compressing the reference images by \textbf{2601$\times$}.}
    \label{fig:reference-compression}
\end{figure}

\ktedit{
However, detecting changes via downsampled reference images will not be as accurate as detecting changes using full-resolution reference images.
Fortunately, under appropriate illumination alignment, if a tile does not change significantly from the reference, it will still have a low difference at a low resolution.
So only changed tiles might be mis-detected as unchanged (\ie {\em false negatives}).
This can happen when the pixel changes in a tile are averaged out when downsampling the new and reference images. 
}

To minimize the false negatives, \name uses a low threshold $\theta$ to detect more changed tiles without misclassifying an unchanged tile as changed.
To demonstrate this, Figure~\ref{fig:reference-compression} shows that when detecting changed tiles under different resolutions, we can always choose a threshold $\theta$ such that about 40\% of tiles are labeled as changed with almost no unchanged tiles being misclassified (\ie they are actually unchanged tiles when seen at a higher resolution).
For instance, when an image is downsampled by over 2600$\times$, only 1.7\% of tiles are mislabeled.
In our experiments, we choose a static threshold $\theta$ based on the data in the first year and evaluate it on the data in the second year.

\mypara{Only uploading changed areas}
As \name applies reference-based encoding, which does not encode the unchanged areas in the captured satellite imagery, this saves the on-board storage space used for storing captured imagery by about 80\% (as 80\% of the areas do not need to be encoded on average, as shown in \S\ref{sec:evaluation}) and enables \name to use the following optimization to further reduce the usage of uplink.
Concretely, \name locally \textit{caches} the reference images onboard the satellite for all locations the satellite will visit and only uploads \textit{changed areas} when uploading a new reference image to the satellite.
The overhead of such caching is marginal (about 9\%, detailed estimation in \S\ref{appendix:storage-overhead} compared to the existing storage space used to store all reference images on-board),
    and thus fits into the storage space conserved by reference-based encoding.
Also, caching reference images on-board allows \name to handle occasional uplink disconnection (more details in \S\ref{sec:implementation}).





\mypara{Uploading only cloud-free areas}
\name requires \textit{cloud-free} reference images to detect terrestrial changes.
However, accurately identifying cloud-free imagery can be computationally expensive, as it requires sequences of images as the input and tens of layers of the neural network to accurately detect light haze and faint clouds~\cite{see-through-cloud}.
As a result, the satellite does not have sufficient computation on board to accurately detect cloud-free imagery.

\name takes an alternative design that identifies cloud-free imagery by letting the ground stations \textit{re-detect} the cloud in the imagery downloaded from satellites using a more compute-intensive but more accurate cloud detector.
This allows \name to accurately detect cloud-free images without incurring extra computation costs on the satellite.

\tightsection{Implementation}
\label{sec:implementation}

\begin{figure}[t!]
\centering
    \includegraphics[width=.7\columnwidth]{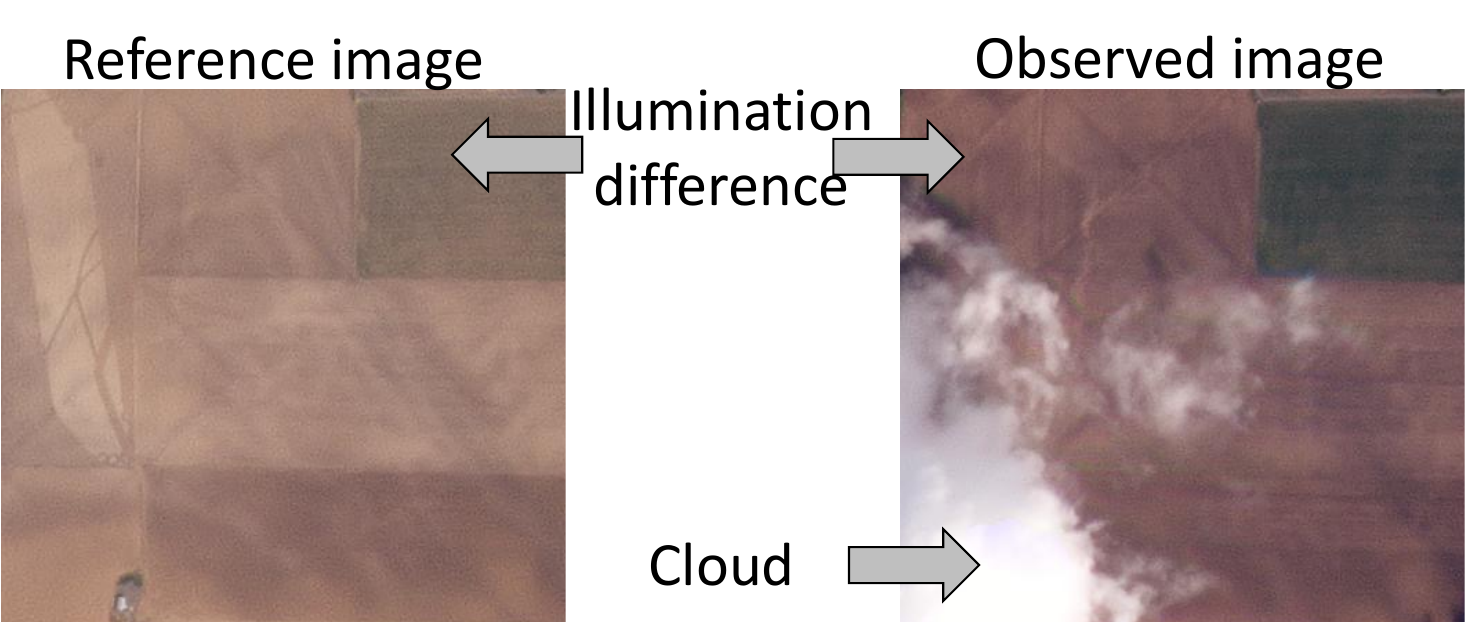}
    \vspace{-0.3cm}
    \tightcaption{Two satellite images captured consecutively do have similar content on the ground, but the pixel difference can still be significant due to cloud and illumination differences. Image © 2023 Planet Labs PBC.}
    \label{fig:satellite-imagery-dynamism}
\end{figure}



\mypara{Illumination and cloud}
In satellite imagery, the time gap between two consecutively-captured images can be hours~\cite{planet} or days~\cite{sentinel-2}.
As a result, two consecutive images in the image sequence can differ a lot in terms of pixel values due to the illumination condition and the cloud condition difference (as shown in Figure~\ref{fig:satellite-imagery-dynamism}), making the general-purpose change detector (like the change detector in videos~\cite{h264,h265,reducto,glimpse}) no longer suitable for satellite imagery compression.

Note that there are other potential sources (\textit{e.g.} sensor noise, image misalignment) that can also trigger large pixel differences. \name does not explicitly address them, as they only appear in raw data sensed by the satellite, which is not accessible in public datasets.

\mypara{On-board change detector}
To be robust to different cloud and illumination conditions, \name employs the following change detection workflow:
\begin{packeditemize}
    \item Cloud removal: detect highly cloudy areas in the satellite imagery using a decision tree classifier, and remove this part of the data (by filling the corresponding pixels with zero pixel value).
    \item Image dropping: drop those images with cloud coverage greater than 50\%.
    \item Illumination alignment: align the illumination between the reference image and the captured image on less-cloudy areas using standard linear regression (since the illumination condition affects the pixel value linearly~\cite{illumination-model}).
    \item The difference detection process is mentioned in \S\ref{subsec:reference-compression}.
    \item Encoding: encode the changed tiles.
\end{packeditemize}

Note that the abovementioned pipeline drops those images that are highly cloudy.
Acknowledging that this makes \name less applicable in cloud-sensitive applications like weather forecasting, we argue that a wide range of applications (\eg autonomous road detection, precision agriculture, etc) treat highly cloudy data as of low value and will typically filter them out as they are focusing on the geographical content on the ground rather than the cloud.
Prior work also makes similar observations~\cite{kodan,satroi}.

\mypara{Encoding changed tiles}
\name encodes those changed tiles by selecting the changed tiles as region-of-interest and runs region-of-interest encoding on the whole image using an off-the-shelf JPEG-2000 encoder (Kakadu~\cite{kakadu}).
While encoding such images, \name makes sure that the bit spent on each encoded tile is a constant $\gamma$ by configuring the bit-per-pixel parameter of the Kakadu encoder as $\gamma$ times the percentage of tiles that are changed.

\mypara{Choosing parameters for \name}
\name introduces two parameters: change detection threshold $\theta$ (\S\ref{subsec:reference-compression}) and bit-per-pixel $\gamma$ for each downloaded imagery.
\name chooses $\theta$ by profiling last year's data on one single location, and uses this parameter on this year's data for all locations.
\name then varies $\gamma$ to trade-off between the downlink usage and the imagery quality.

\mypara{Fitting into on-board computation constraint}
In the change detector of \name, both illumination alignment and difference detector can be cheaply done on-board.
However, accurate cloud detection can be compute-intensive, as it typically requires using a series of images as input and tens of layers of neural network to accurately detect thin clouds~\cite{see-through-cloud}.

To reduce the computation cost of the on-board cloud detector, the main observation of \name is that it is tolerable if the cloud detector fails to detect some clouds (as they will be detected as changed and downloaded to the ground, which is tolerable as long as the downlink saving of \name is still sizable), but it is harmful if the cloud detector wrongfully detect non-cloudy area as cloudy as it will trigger \name to discard these areas, where there may be changes.
Using this observation, \name only detects those easy-to-detect clouds (\eg heavy clouds), which can be done by a cheap decision-tree-based detector (as the temperature of heavy clouds significantly differs from the nearby ground and can be easily detected using the InfraRed band) while still ensuring that almost all areas that the cloud detector detects are cloudy.
Further, \name also detects the cloud under a downsampled version of the captured imagery ($64\times$, width and height) as \name only uses the cloud detection to identify which 64$\times$64 tiles need to be downloaded.

After these optimization techniques, \name can make  cloud detection faster compared to running JPEG-2000 encoding, while making sure that over 99\% of areas detected are actually cloudy.

\mypara{Handling different bands}
Unlike traditional RGB images, satellite imagery typically has multiple bands (\eg near-infrared band and short-wave infrared band) and the amount of changes of different bands (B1 -- B12) on cloud-free areas are different. 
This is because some of the bands aim to monitor the air quality and thus do not change significantly in cloud-free areas (\eg B9), while some of the bands change a lot due to their temperature sensitivity (\eg vegetation bands such as B7, B8, and B8a measures the concentration of chlorophyll, which is sensitive to temperature).
To handle such heterogeneity between bands, \name treats each band separately, which means that \name detects changes band-by-band and updates the reference images band-by-band, allowing \name to mark different areas as changed and download different amounts of changes for different bands.

\mypara{Handling bandwidth fluctuation}
We introduce how \name handles the fluctuation of uplink and downlink bandwidth.

We first handle uplink alteration. As \name locally caches the reference images, when the capacity of the uplink alters, \name can skip the updating of some reference images (the skipped images are randomly chosen) and rely on the old reference images cached on-board for change detection, with a slight increase on the downlink capacity.

We then handle downlink alteration. \name relies on the \textit{layered codec} to handle the alteration of downlink.
    The characteristic of layered codec is that by encoding the images to multiple layers, \name can smoothly trade-off between downlink bandwidth and the quality of downloaded imagery quality: the ground can download more layers to receive high-quality imagery when having sufficient downlink bandwidth or download fewer layers to the ground when the downlink bandwidth is limited.
    The feature of layered codec is widely supported by existing imagery encoders on the satellite (\eg JPEG-2000 encoders~\cite{openjpeg,kakadu}).
    We note that \name downloads the same amount of layers for all locations that \name will download during one ground contact.

\mypara{Guaranteed downloading}
As some changed tiles will not be downloaded to the ground by \name (\S\ref{subsec:reference-compression}), \name performs guaranteed downloading that downloads cloud-free imagery in its entirety at a low frequency (once every month).

\tightsection{Evaluation}
\label{sec:evaluation}

In this section, we pick two state-of-the-art satellite imagery compression systems as our baseline and evaluate \name against  on two satellite imagery datasets.
The key takeaway of our evaluation is three-fold:
\begin{packeditemize}
    \item Compared to the state-of-the-art onboard compression schemes, \name reduces the downlink bandwidth usage by 1.3-3.3$\times$ without hurting the imagery quality on all bands.
    \item These improvements are achieved without using more uplink bandwidth than currently available, or more compute or storage resources than the baselines.
    \item With more satellites in a constellation, \name can further reduce the amount of downlink bandwidth usage.
\end{packeditemize}

\tightsubsection{Experimental setup}

\begin{table}[]
\footnotesize
\centering
\begin{tabular}{|c|c|c|}
\hline
Section                                                                                & Properties                 & Values                                                  \\ \hline
\multirow{4}{*}{\begin{tabular}[c]{@{}c@{}}Connectivity\end{tabular}} & Ground contact duration    & \textit{10 minutes}~\cite{leoconn,ground-contact-time}  \\ \cline{2-3} 
                                                                                       & Ground contact per day     & \textit{7 times}~\cite{ground-contact-time}             \\ \cline{2-3} 
                                                                                       & Uplink bandwidth           & 250 kbps~\cite{dove-uplink-downlink}                    \\ \cline{2-3} 
                                                                                       & Downlink bandwidth         & 200 Mbps~\cite{dove-uplink-downlink}                    \\ \hline
Hardware    & On-board storage           & \textit{360 GB}~\cite{superdove-storage}                      \\ \hline
\multirow{4}{*}{\begin{tabular}[c]{@{}c@{}}Image\end{tabular}}        & Image resolution           & 6600$\times$4400~\cite{planet-ground-sampling-distance} \\ \cline{2-3} 
                                                                                       & Image channels             & RGB + InfraRed~\cite{planet-ground-sampling-distance}   \\ \cline{2-3} 
                                                                                       & Raw image file size            & \textit{150 MB}~\cite{planet-image-size}               \\ \cline{2-3} 
                                                                                       & Ground sampling distance   & 3.7 meters~\cite{planet-ground-sampling-distance}       \\ \hline
\end{tabular}
\tightcaption{Characterizing the specifications of Doves constellation from the year 2017 to the year 2018. Some of the data are not publically available and we infer them from other sources (we \textit{italicized} the data that we inferred, with the corresponding sources cited next to the inferred data).}
\label{tab:hardware-specs}
\end{table}

     

\begin{table*}[]
\footnotesize
\centering
\begin{tabular}{ccccccccc}
\hline
           & Why using this dataset                                                                                          & \begin{tabular}[c]{@{}c@{}}Number of satellites\\ included in our dataset\end{tabular} & Locations & \begin{tabular}[c]{@{}c@{}}Coverage of\\ each location\end{tabular} & \begin{tabular}[c]{@{}c@{}}Ground sampling\\ distance (GSD)\end{tabular} & Duration & \begin{tabular}[c]{@{}c@{}}Number\\ of bands\end{tabular} & \begin{tabular}[c]{@{}c@{}}Cloud\\ coverage\end{tabular} \\ \hline\hline
Planet     & \begin{tabular}[c]{@{}c@{}}Show that \name saves \\ more downlink when\\ there are more satellites\end{tabular} & 48                                                                                      & 1         & 36 km$^2$                                                           & 3.0 - 4.1 m                                                              & 3 months & 4                                                         & <5\%                                                     \\ \hline
Sentinel-2 & \begin{tabular}[c]{@{}c@{}}Test \name on a\\ wide range of content\end{tabular}                                 & 2                                                                                      & 11         & 1600 km$^2$                                                         & 10 m                                                                     & 1 year   & 13                                                        & $\leq$100\%                                                    \\ \hline
\end{tabular}
\tightcaption{The datasets used in our evaluation. One from Planet~\cite{sentinel-2} and the other from Sentinel-2~\cite{planet}.}
\label{tab:dataset}
\end{table*}

\begin{figure*}[t]

    \begin{subfigure}[b]{0.16\textwidth}
         \centering
         \includegraphics[width=\textwidth]{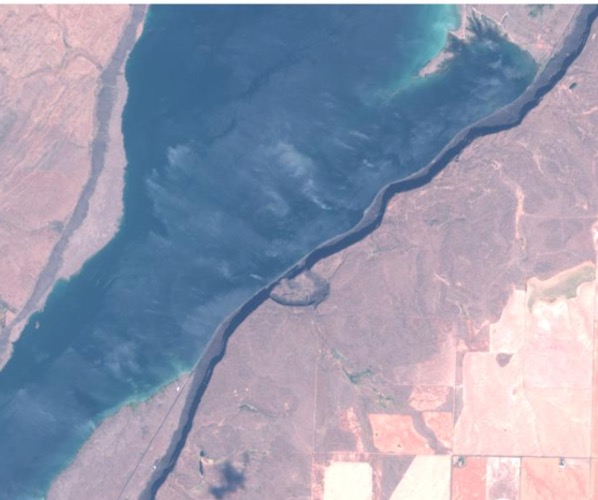}
         \caption{Sentinel-2: River}
     \end{subfigure}
     \begin{subfigure}[b]{0.16\textwidth}
         \centering
         \includegraphics[width=\textwidth]{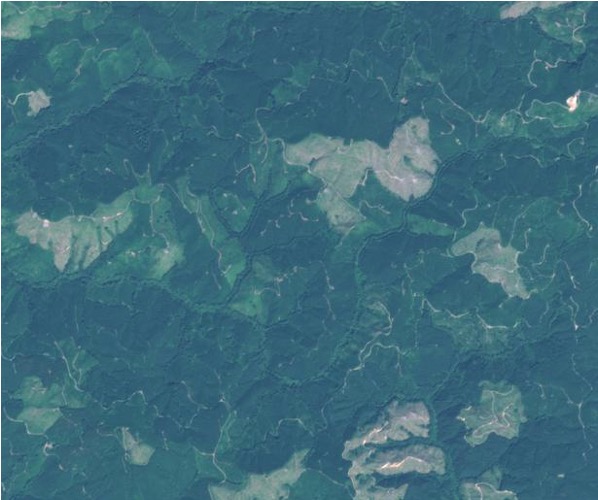}
         \caption{Forest}
     \end{subfigure}
     \begin{subfigure}[b]{0.16\textwidth}
         \centering
         \includegraphics[width=\textwidth]{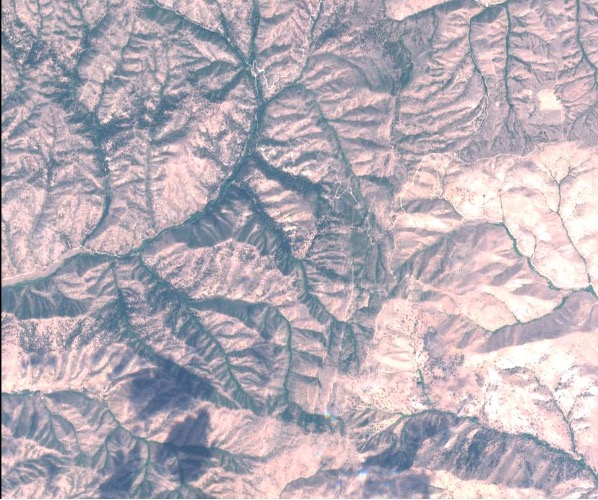}
         \caption{Mountain}
     \end{subfigure}
     \begin{subfigure}[b]{0.16\textwidth}
         \centering
         \includegraphics[width=\textwidth]{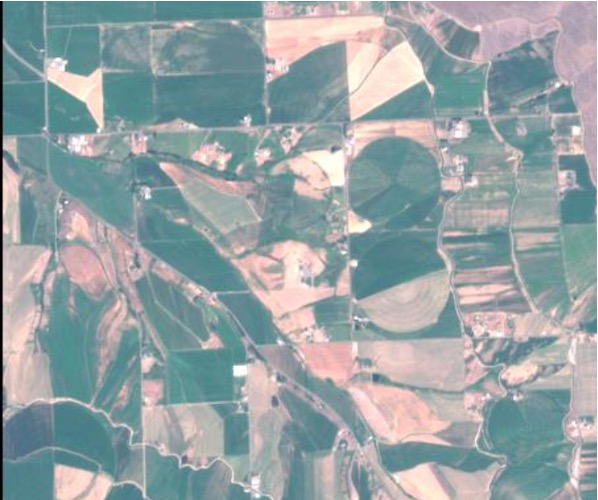}
         \caption{Agriculture}
     \end{subfigure}
     \begin{subfigure}[b]{0.16\textwidth}
         \centering
         \includegraphics[width=\textwidth]{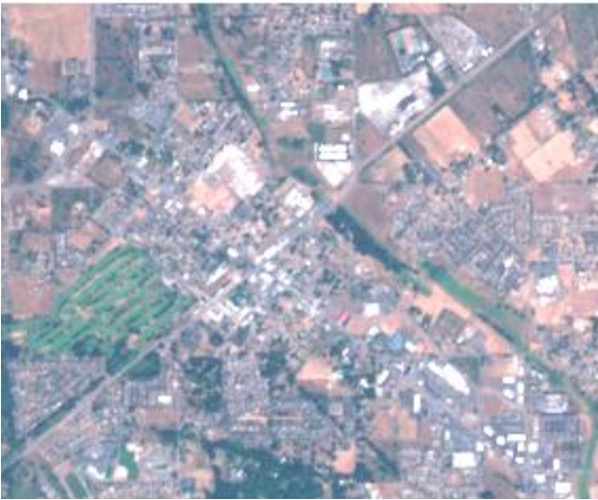}
         \caption{City}
     \end{subfigure}
     \begin{subfigure}[b]{0.16\textwidth}
         \centering
         \includegraphics[width=\textwidth]{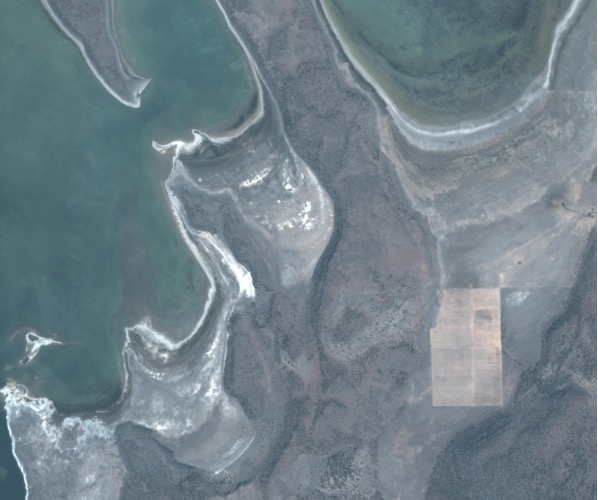}
         \caption{Planet: Coastal}
         \label{fig:dataset-diversity:planet}
     \end{subfigure}
     \vspace{-0.4cm}
     \tightcaption{Snapshots of images at sampled locations in our testing datasets, which show a variety of geographical content.}
     \label{fig:dataset-diversity}
\end{figure*}

\mypara{Dataset}
We evaluate \name on two datasets (Table~\ref{tab:dataset} illustrates the details of these two datasets):
\begin{packeditemize}
    \item \textbf{\textit{Rich-content dataset}}: we collect 1-year images on 11 geographical locations in Washington State (where each location is of size 1600 km$^2$) from Sentinel-2 dataset~\cite{sentinel-2}.
    We sample images from Washington State as it contains a wide variety of geographical contexts, including fluvial landscapes, agricultural areas with varied irrigation systems, mountainous regions with large elevation changes, etc, as shown in Figure~\ref{fig:dataset-diversity}a-e.

    To handle the large volumn of this dataset, we downsample the images in this dataset by 4$\times$, width and height, where we confirmed on one location that such downsampling does not affect the savings of \name.

    However, Sentinel-2 dataset~\cite{sentinel-2} only contains two satellites in its constellation. 
    To further show the potential of \name's constellation-wide change-based encoding, we incorporate another dataset with lower coverage but with more satellites available.
    
    \item \textbf{\textit{Large-constellation dataset}}: we use Planet dataset~\cite{planet} that contains multiple satellites in its constellation to showcase the potential of \name's constellation-wide change-based encoding.
    Due to the quota limit of the Planet dataset, we only sample images on one randomly sampled location in the U.S. (the content is illustrated in Figure~\ref{fig:dataset-diversity:planet}), with cloud coverage smaller than 5\%.
    The dataset we sampled contains 48 satellites in total.
\end{packeditemize}
\mypara{Real-world satellite specifications}
We show the satellite specifications in Table~\ref{tab:hardware-specs}.

\mypara{Modelling uplink and downlink}
We evaluate \name using the uplink and downlink specifications from Doves constellation. Specifically:
\begin{packeditemize}
    \item Uplink: we assume that the uplink is of a constant bandwidth (250 kbps~\cite{dove-uplink-downlink}) and the connection duration is 10 minutes~\cite{leoconn,ground-contact-time}. 
    Here we assume that the uplink bandwidth is a constant, as the uplink leverages the S-band to communicate~\cite{Sentinel-2-uplink}, which is of low frequency, and thus severe weather conditions do not significantly affect its bandwidth~\cite{s-band-rain}.
    \item Downlink: we assume that the ground contact duration is 10 minutes~\cite{leoconn,ground-contact-time} and calculate the minimum average bandwidth required to download a fixed amount of images.
\end{packeditemize}


\mypara{Metrics}
\name aims to reduce the downlink demand without hurting the quality of downloaded images.
We measure the required downlink bandwidth by dividing the amount of data needed to be streamed during one ground contact by the ground contact time (10 minutes~\cite{leoconn,ground-contact-time}) and measure the image quality by using Peak Signal-to-Noise Ratio (PSNR for short).
This aligns with satellite compression literature~\cite{indradjad2019comparison,gunasheela2018satellite,faria2012performance,shihab2017enhancement}.
Further, prior work shows that a higher Peak Signal-to-Noise Ratio typically leads to higher application-side performance~\cite{psnr-38-classification,psnr-38-high}.


\mypara{Baselines}
We consider two state-of-the-art baselines for on-board satellite imagery compression:
\begin{packeditemize}
    \item {\bf Kodan}~\cite{kodan}: drop low-value cloud data and download remaining non-cloudy areas.
    \item {\bf SatRoI}~\cite{satroi}: running reference-based encoding using a fixed reference image.
\end{packeditemize}
We evaluate \name on the standard JPEG-2000 image encoder, which is already used by existing satellites~\cite{planet-jpeg-2000,sentinel-2-jpeg-2000}.
While there is a line of work on single satellite imagery encoders (\eg augmenting traditional image codecs~\cite{edge-prediction,progressive-jpeg2000,CALIC,linear-prediction,band-reordering,band-reordering-segmentation,band-reordering-prediction,band-prediction} or developing neural-based codecs such as autoencoders~\cite{autoencoder-1,autoencoder-2,autoencoder-3,autoencoder-4,autoencoder-5}), \name is in parallel with this line of work (as \name focuses on leveraging the redundancy \textit{between} images for further compression) and one can use \name on top of existing single satellite imagery encoders.


\tightsubsection{Experimental results}

\begin{figure}
    \centering
    \begin{subfigure}[b]{0.50\columnwidth}
         \centering
         \includegraphics[width=\textwidth]{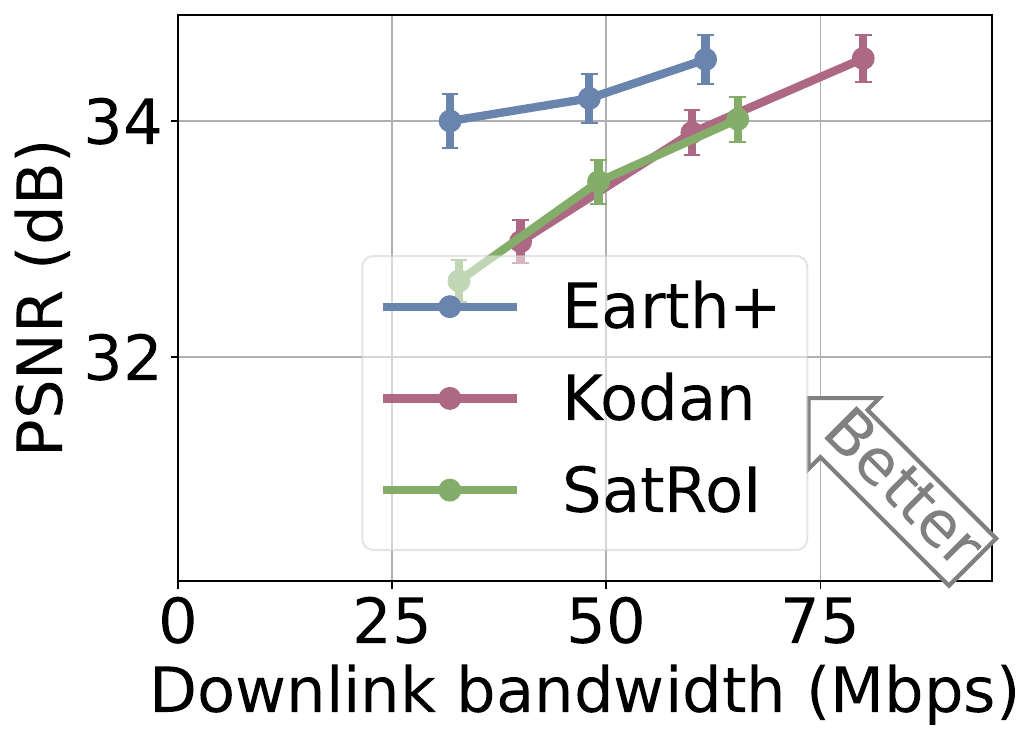}
         \caption{Sentinel-2 dataset}
         \label{fig:psnr-bw:sentinel}
     \end{subfigure}
    \begin{subfigure}[b]{0.48\columnwidth}
         \centering
         \includegraphics[width=\textwidth]{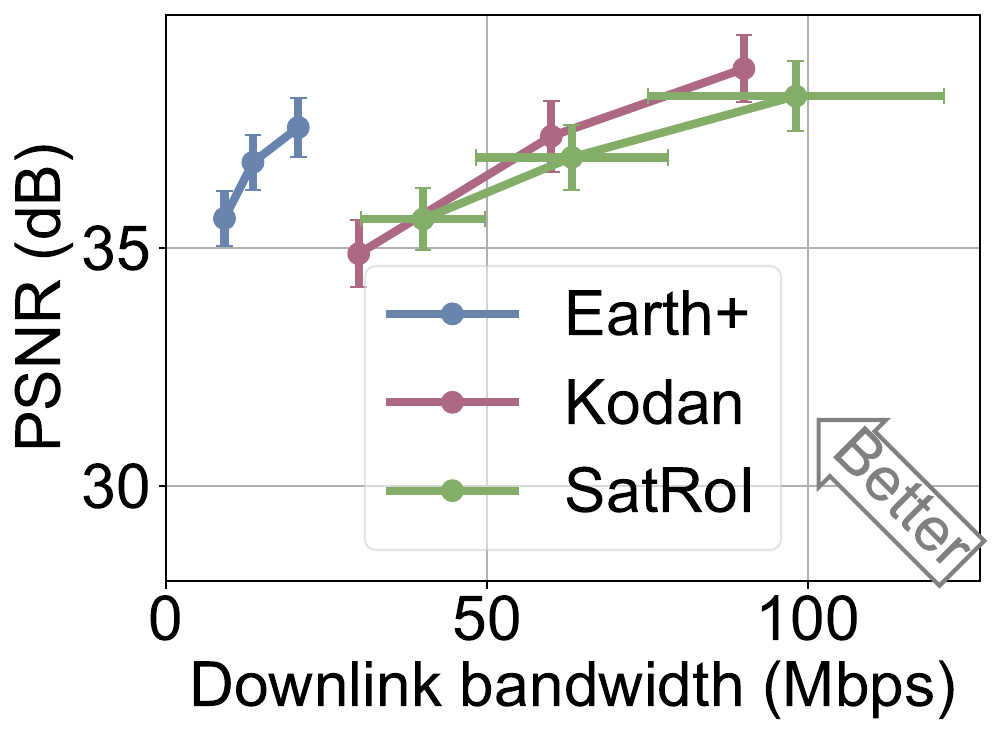}
         \caption{Planet dataset}
         \label{fig:psnr-bw:planet}
     \end{subfigure}
    \vspace{-0.3cm}
    \tightcaption{
    \kt{Redo sentinel-2 results}
    \name requires less downlink bandwidth without sacrificing the image quality (measured in PSNR). The vertical and horizontal bar shows the standard deviation of the mean (some of the horizontal bars are occluded as they are too short).
    }
    \label{fig:psnr-bw}
\end{figure}

\begin{figure}
    \centering
    \includegraphics[width=0.7\columnwidth]{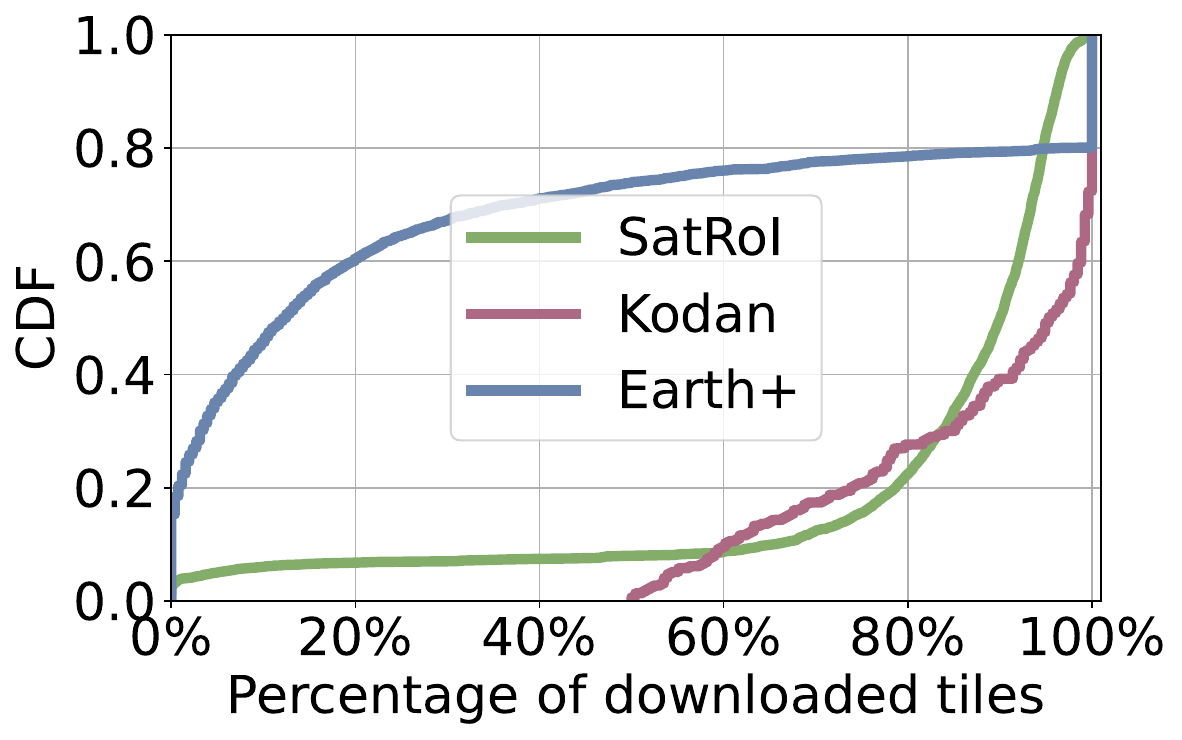}
    \includegraphics[width=0.7\columnwidth]{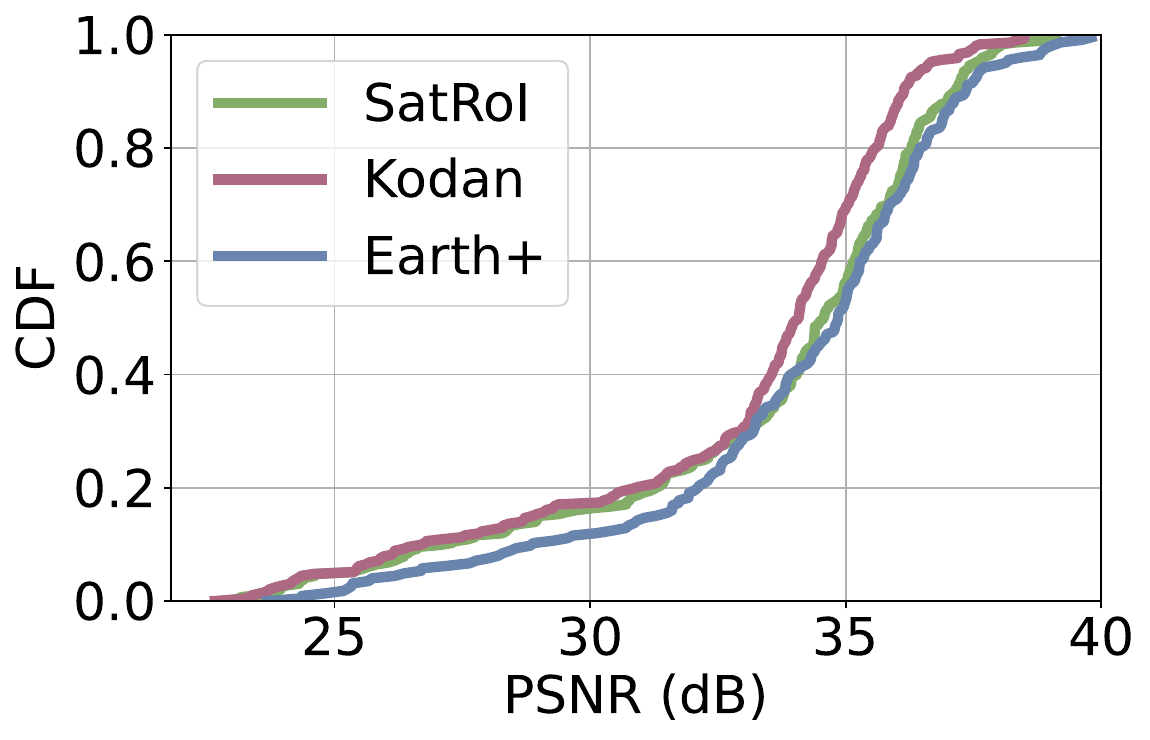}
    \tightcaption{The CDF of the percentage of downloaded tiles and the PSNR of \name and baselines. We can see that \name downloads much fewer tiles compared to the baselines while achieving higher PSNR.
    }
    \label{fig:change-cdf}
\end{figure}


\begin{figure}
    \centering
    \includegraphics[width=0.7\columnwidth]{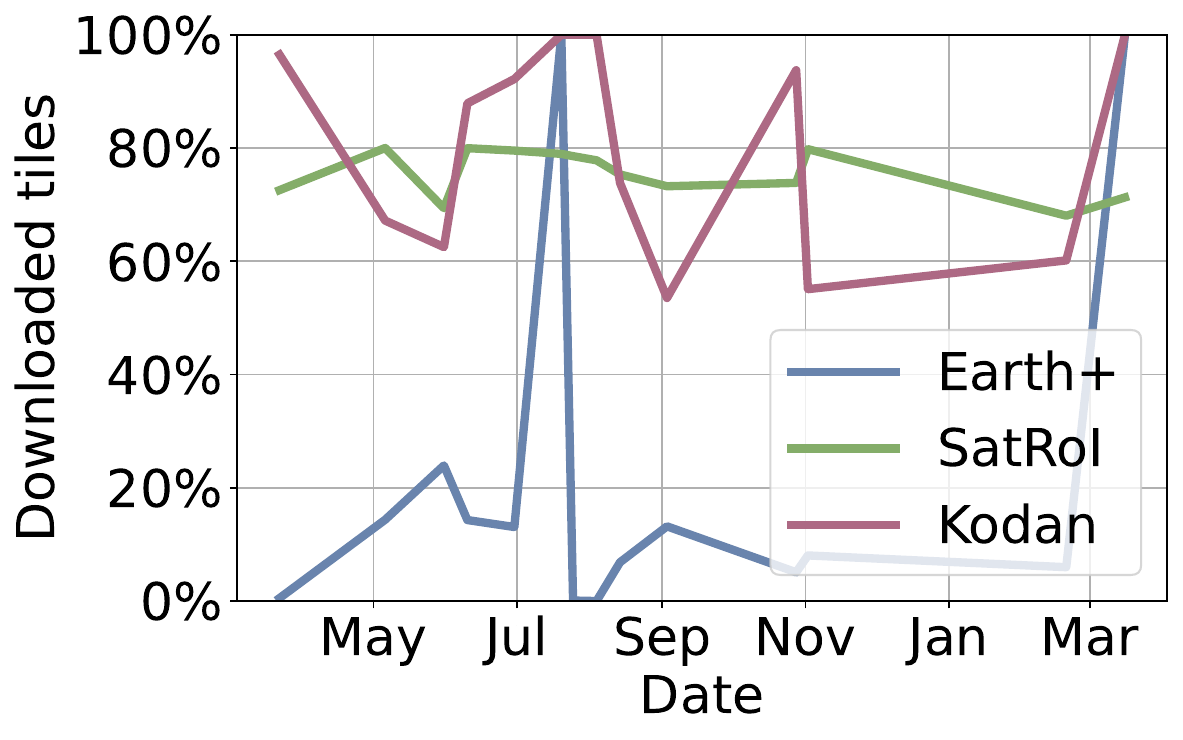}
    \includegraphics[width=0.7\columnwidth]{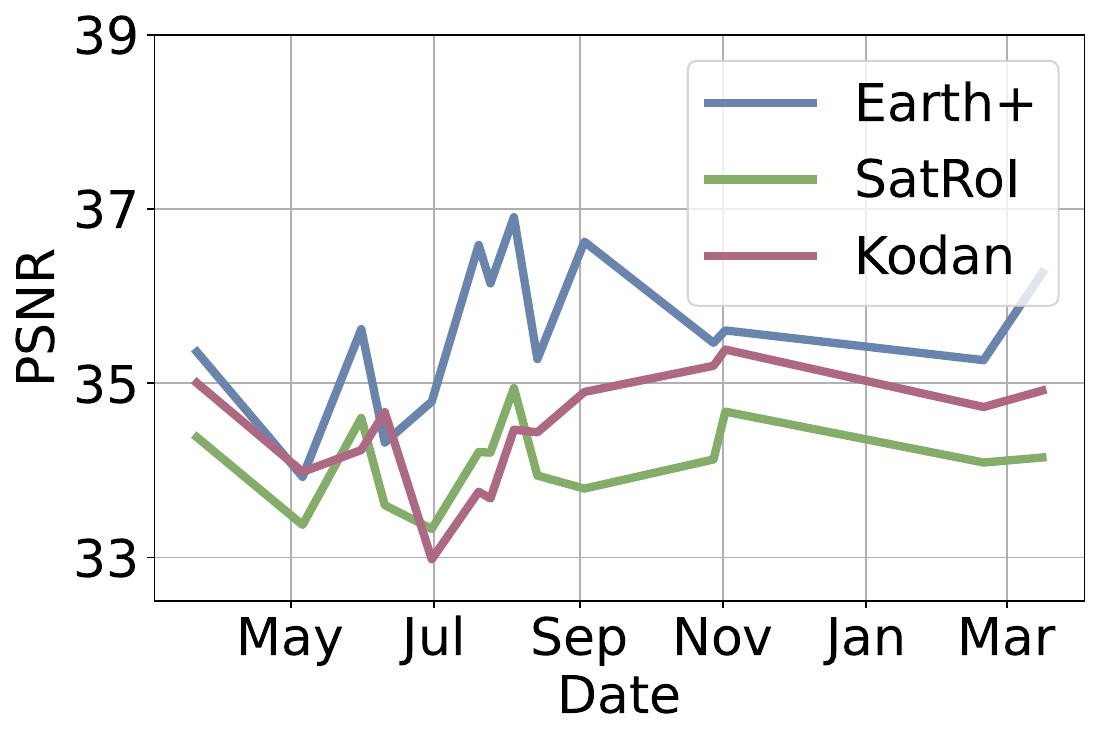}
    \tightcaption{The time series of the percentage of downloaded tiles and the PSNR of \name and baselines on one location.
    }
    \label{fig:change-time-series}
\end{figure}

\mypara{Saving downlink bandwidth without hurting PSNR}
We show that \name has a better PSNR--downlink demand trade-off compared to the baselines.
As shown in Figure~\ref{fig:psnr-bw:sentinel}, \name saves 1.3-2.0$\times$ downlink bandwidth without hurting the PSNR of the images compared to the strongest baseline in the Sentinel-2 dataset.
\name further saves downlink bandwidth by 2.8-3.3$\times$ in the Planet dataset, as shown in Figure~\ref{fig:psnr-bw:planet}.
The reason behind such improvement is that \name uses \textit{fresh} reference images to detect and encode changes in the imagery, while Kodan needs to encode all non-cloudy regions (including those regions that are not changed), and SatRoI uses a fixed reference image that nearly detects all regions as changed.
Also, \name has more savings on the Planet dataset. This is because the Planet dataset contains more than 40 satellites, while Sentinel-2 only contains two satellites, and \name's constellation-wide encoding scheme can 
This is because the Planet dataset has more satellites than the Sentinel-2 dataset, and \name's constellation-wide design can benefit from more satellites.

Figure~\ref{fig:change-cdf} further shows the cumulative distribution of the percentage of downloaded tiles and the PSNR of \name and baselines. 
For more than 60\% of the images, \name downloads less than 20\% tiles.
In contrast, the baseline needs to download more than 80\% of tiles for over 70\% of images.
This is because \name can effectively detect and download only those changed tiles, while Kodan may download those cloud-free but unchanged tiles, and SatRoI cannot effectively detect changes.
Kodan needs to download all the non-cloudy areas, and SatRoI may detect and download a lot of changed areas because it uses an older reference frame.
In contrast, \name only downloads the detected changes using a fresher reference image.
There are 20\% of images fully downloaded by \name because of the guaranteed downloading mechanism (\S\ref{sec:implementation}).

Figure~\ref{fig:change-time-series} shows an illustrative timeseries of the downloading behavior of \name and the baselines for one year. 
\name downloads 5-10$\times$ fewer areas than the baselines most of the time.
It only occasionally performs the guaranteed downloading that downloads the full image to keep a high image quality.

    
    

\begin{figure}
    \centering
    \includegraphics[width=0.9\columnwidth]{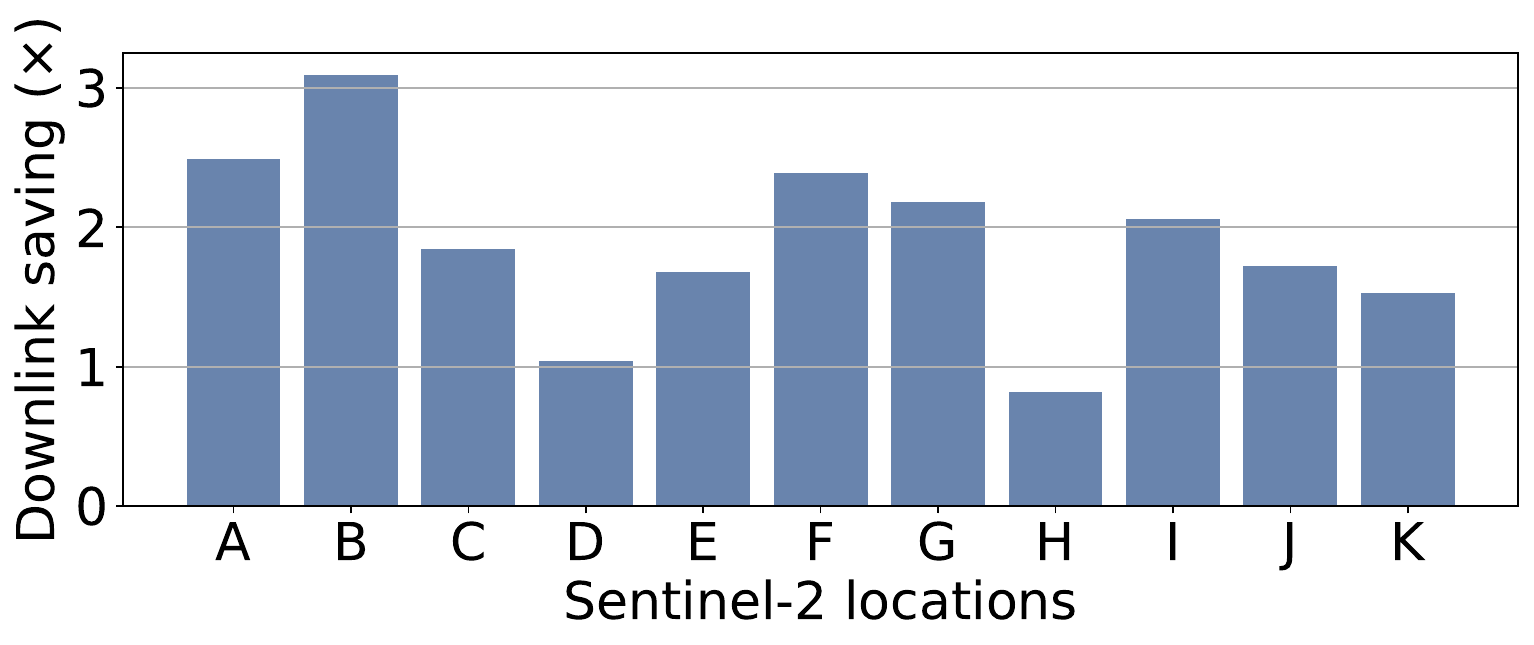}
    \includegraphics[width=0.9\columnwidth]{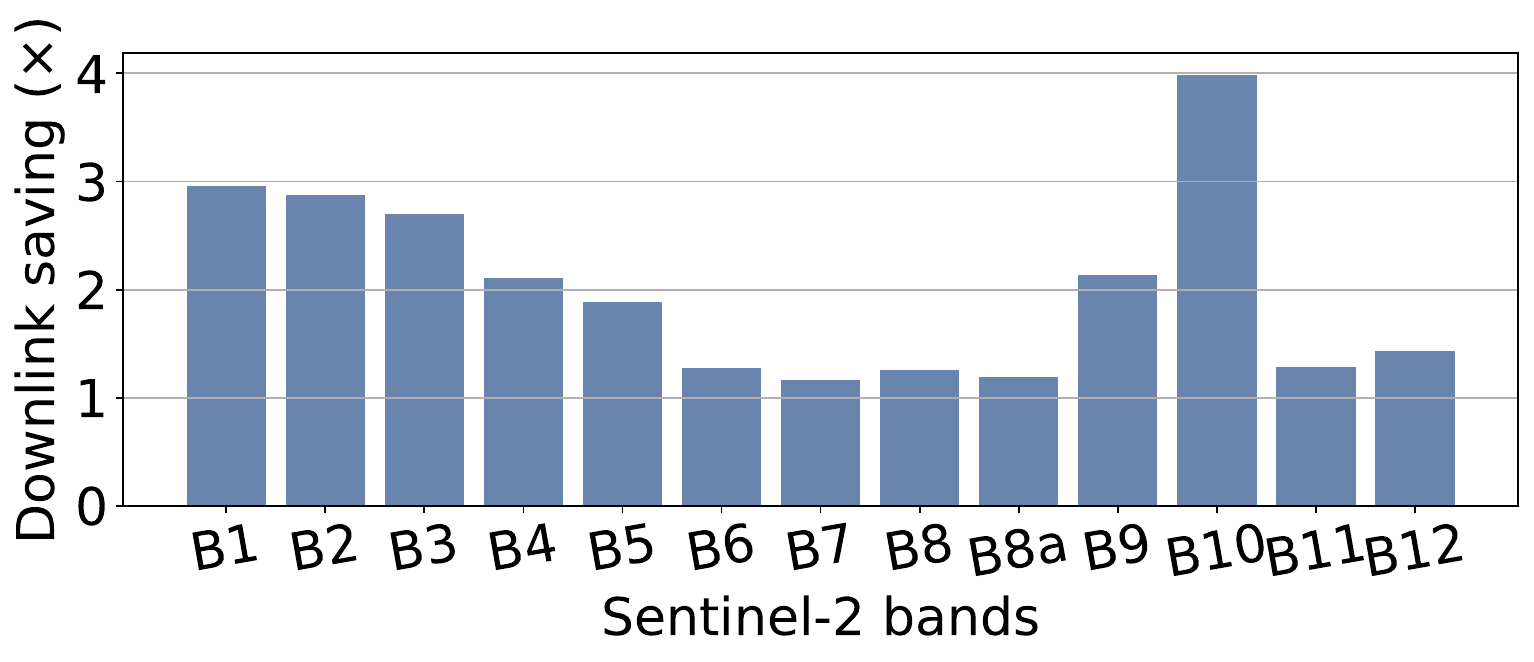}
    \vspace{-0.3cm}
    \tightcaption{The improvement of \name across different locations and bands in the Sentinel-2 dataset. 
    }
    \label{fig:band-improvement}
    \label{fig:group-by-area}
\end{figure}

\mypara{Downlink saving across different locations}
We then calculated \name's saving on the downlink (defined as the strongest baseline with lower PSNR than \name, divided by the downlink usage of \name) grouped by 11 different locations in Washington State.
Figure~\ref{fig:group-by-area} shows that \name is better than the strongest baseline at 10 out of 11 locations. 
However, \name does not improve on H and has only marginal improvement on D, as these two locations are highly snowy during winter and spring, and snow albedo (\ie the reflectance of snow) is constantly changing (\eg old snow has a lower albedo than fresh snow, and dirty snow has a lower albedo than clean snow).
Thus, \name tends always to detect changes and download those tiles that contain snow, lowering the improvement of \name.

\mypara{Downlink saving across different bands}
We further evaluate the downlink saving of \name across different bands of the images (\ie the downlink usage of the strongest baseline with lower PSNR than \name, divided by the downlink usage of \name).
We show the bandwidth saving of \name compared to the strongest baseline. 
As shown in Figure~\ref{fig:band-improvement}, \name can improve on all 13 bands in Sentinel-2 images.
We also observe that the improvement of \name varies between bands.
We show that \name can significantly improve on ground-related bands (\eg RGB bands B2-4) but less significantly on bands that observe the air (\eg water vapor bands).

\begin{figure}
    \centering
    \includegraphics[width=0.8\columnwidth]{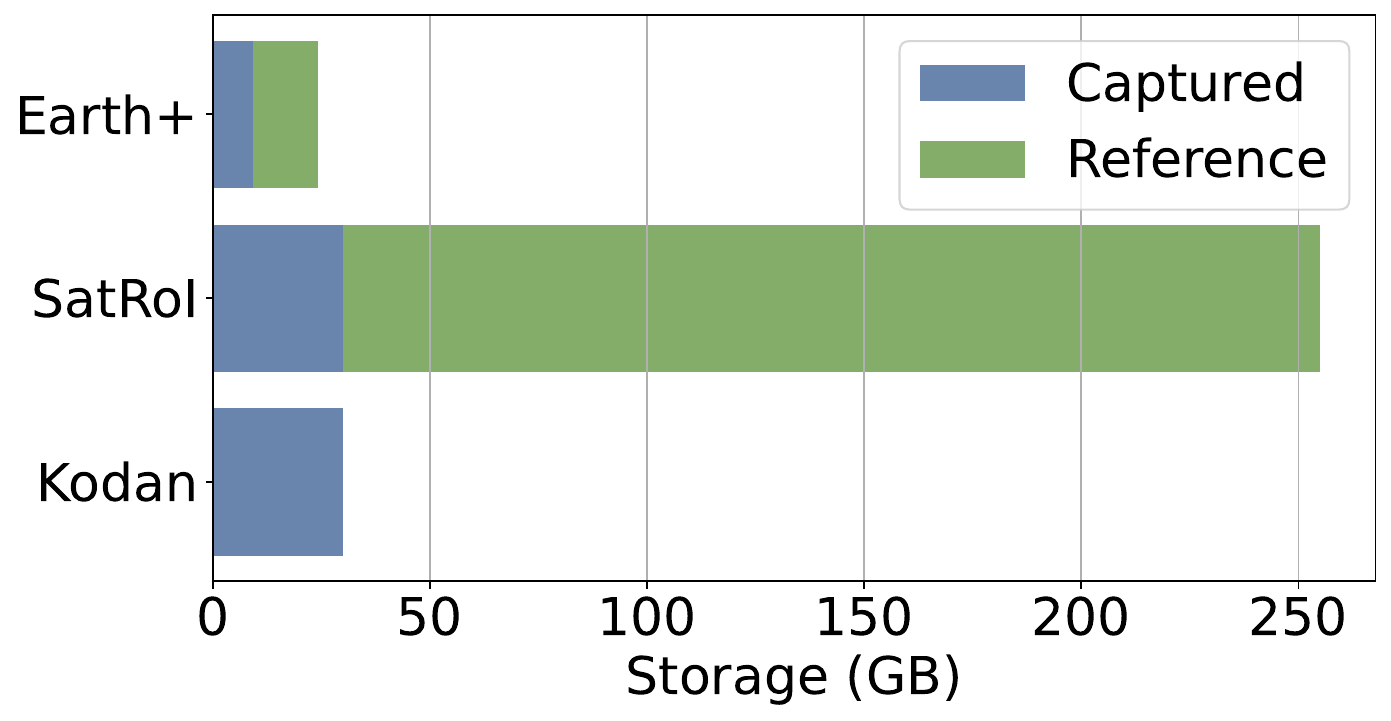}
    \vspace{-0.3cm}
    \tightcaption{Breaking down the storage usage of \name and the baselines. \name saves changed areas on-board rather than whole images like Kodan, which leaves some space for storing reference images.
    }
    \label{fig:storage-usage}
\end{figure}

\begin{figure}
    \centering
    \includegraphics[width=0.85\columnwidth]{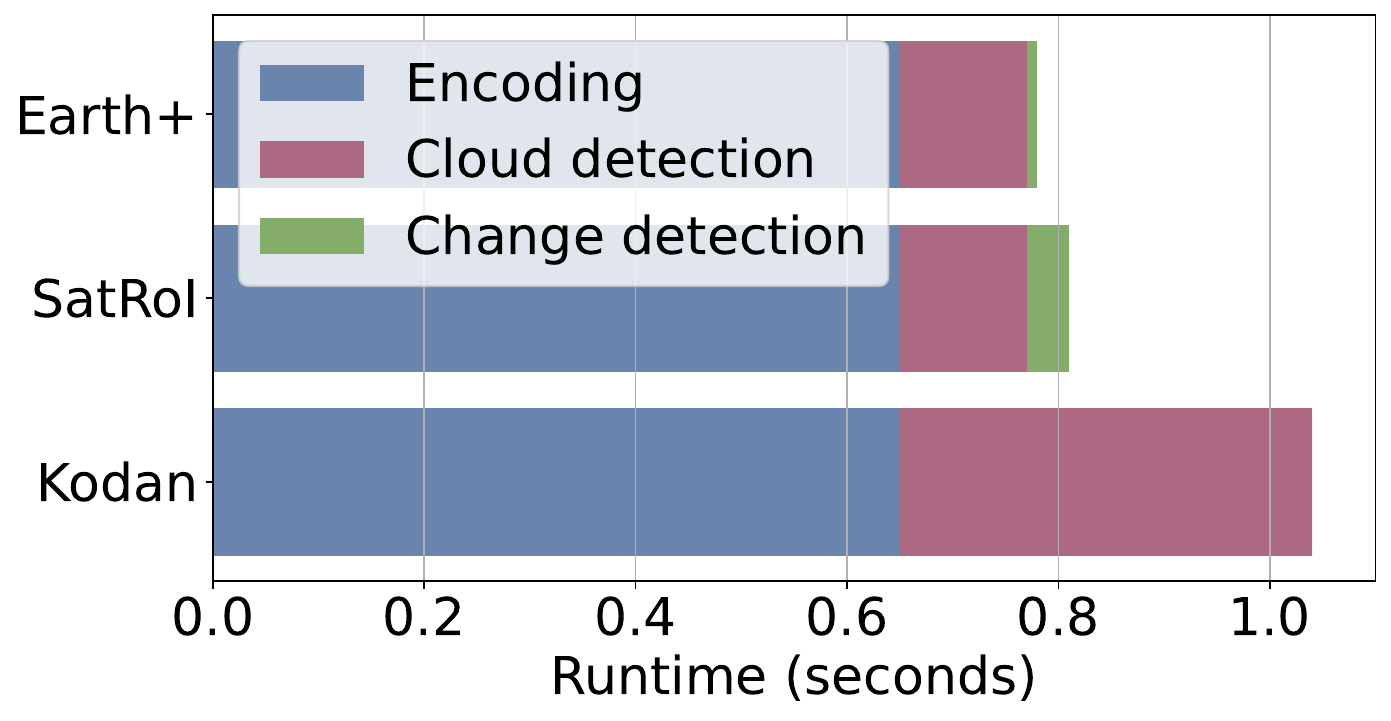}
    \vspace{-0.3cm}
    \tightcaption{Benchmarking the runtime of \name and baselines to process one single satellite imagery. The runtime of \name is lower than the baselines.
    }
    \label{fig:computation}
\end{figure}

\begin{figure}
    \centering
    \includegraphics[width=0.95\columnwidth]{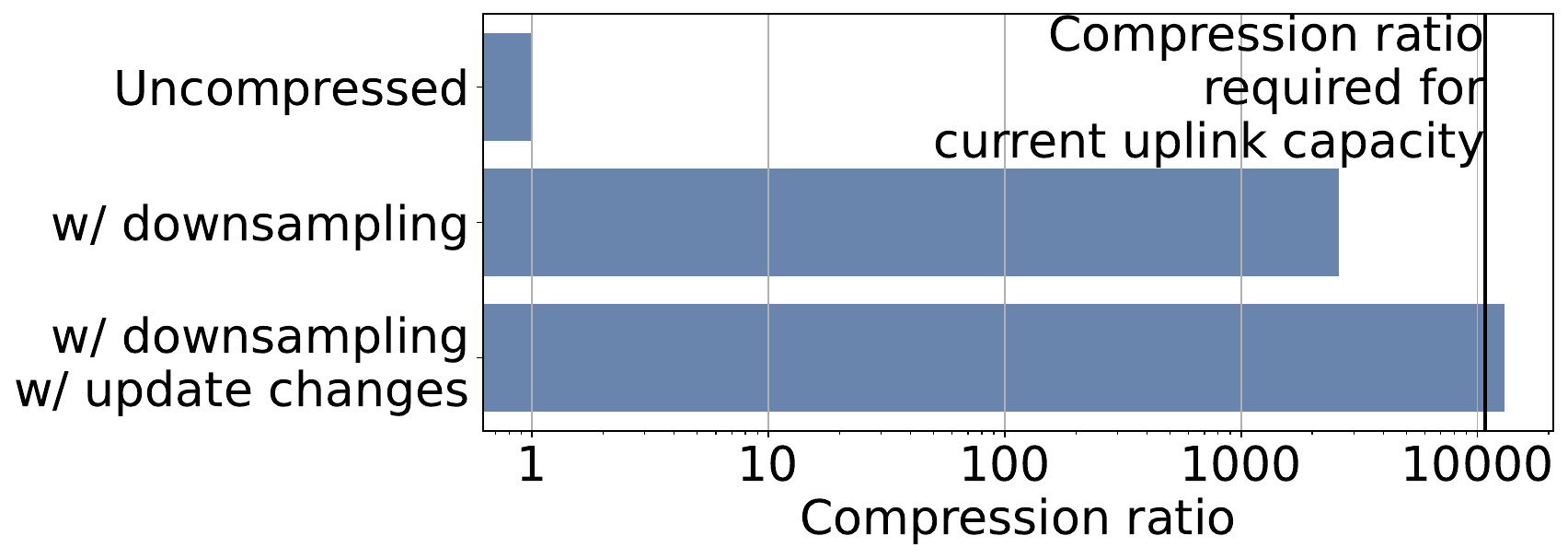}
    \vspace{-0.3cm}
    \tightcaption{\name compresses the reference image by over 10,000$\times$ so that they can fit in the limited uplink capacity, indicated by the black vertical line.
    }
    \label{fig:uplink}
\end{figure}

\begin{figure}
    \centering
    \includegraphics[width=0.9\columnwidth]{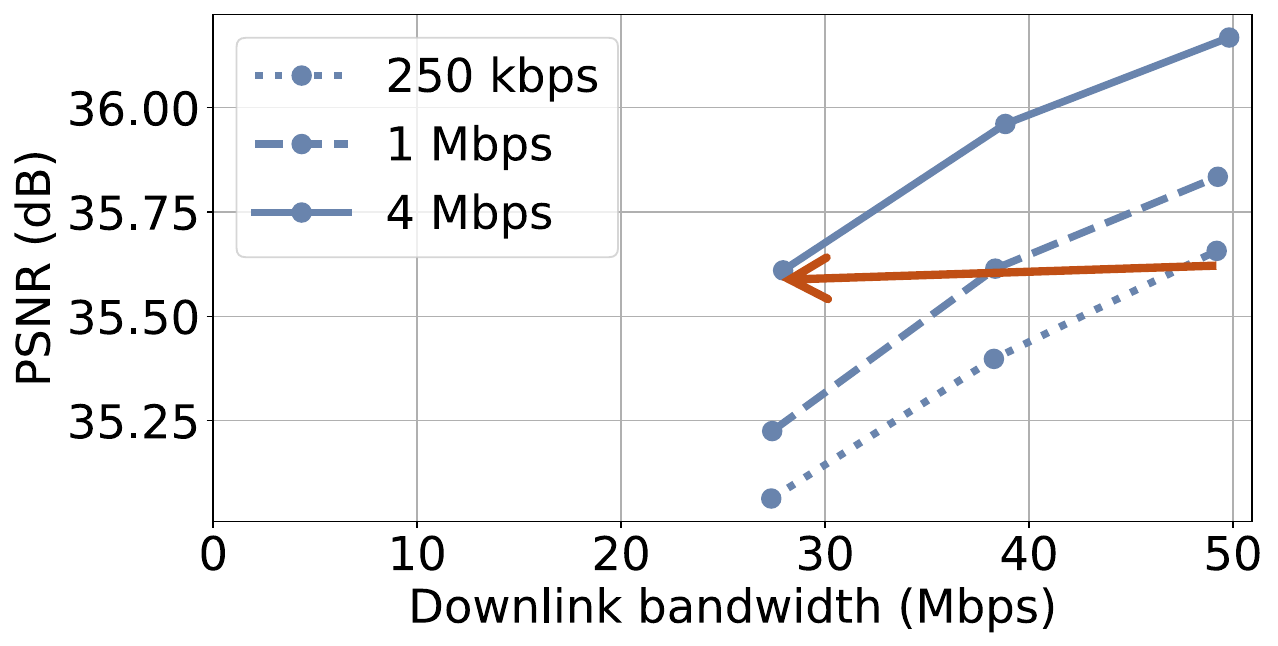}
    \vspace{-0.3cm}
    \tightcaption{
    \name can further reduce the downlink bandwidth usage by 22 Mbps by increasing the uplink bandwidth to 4 Mbps (as indicated by the red arrow).
    }
    \label{fig:uplink-sensitivity}
\end{figure}

\begin{figure}
    \centering
    \includegraphics[width=0.98\columnwidth]{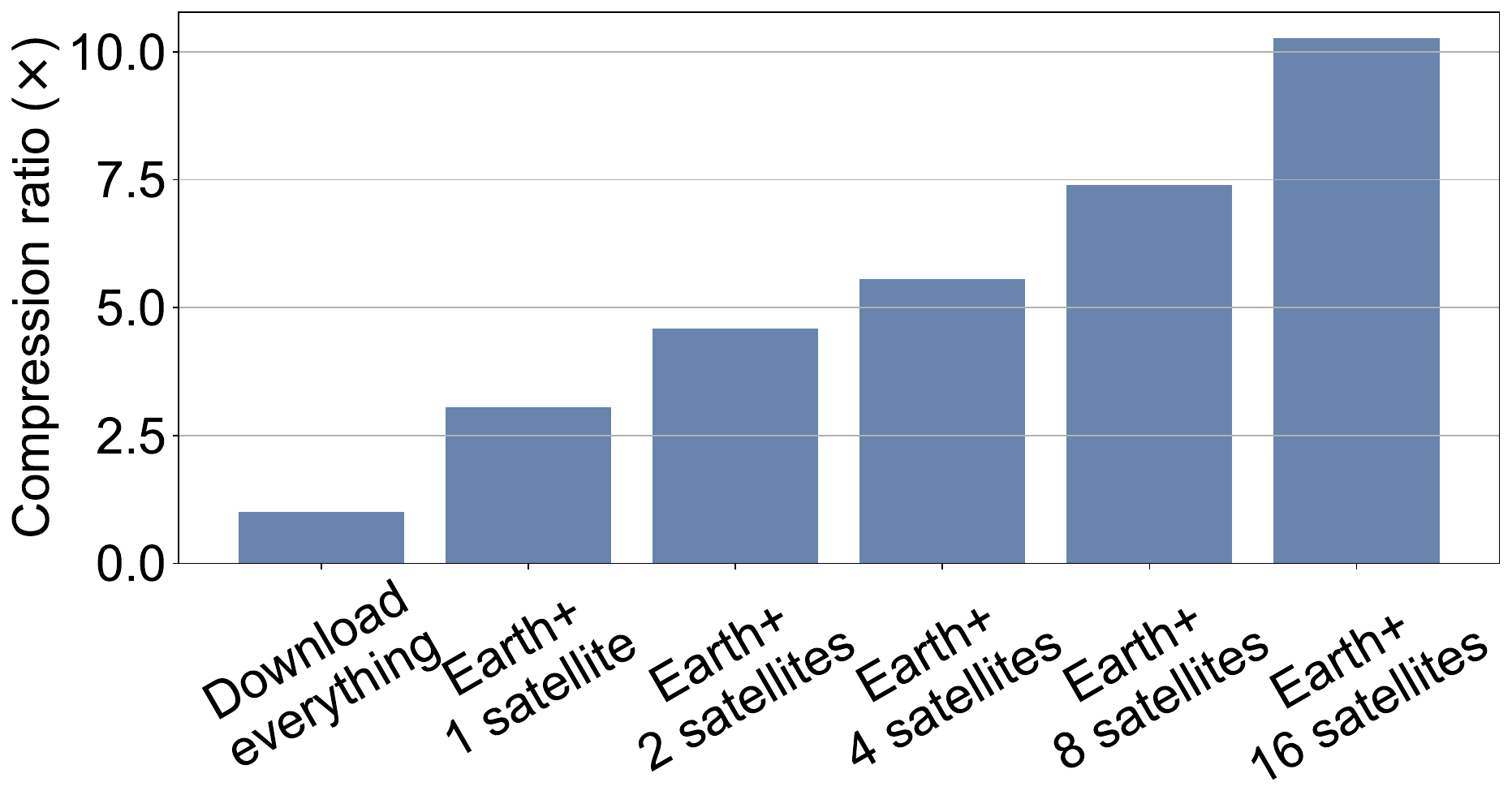}
    \tightcaption{
    \name can compress the images with a higher compression ratio when more satellites are in the constellation.
    }
    \label{fig:more-satellite}
\end{figure}

\mypara{Storage overhead} 
Based on Dove~\cite{dove-constellation} constellation specification, we calculate the storage requirements for \name and two baselines.
As shown in Figure~\ref{fig:storage-usage}, the onboard storage requirements are 30~GB, 255~GB, and 24~GB for SatRoI, Kodan, and \name, respectively.
We note that \name spends less storage space to store the captured imagery, as \name only stores the \textit{changed tiles} on-board.
Further, \name needs much less storage space for reference images, as \name downsamples the reference images to low resolution.
This is because \name stores only changed areas on-board, thus squeezing out storage space to store reference images.

\mypara{Uplink overhead} 
We then evaluate the effectiveness of techniques proposed in \S\ref{subsec:reference-compression} that reduce the update bandwidth usage based on the Sentinel-2 dataset and Dove specification.
Figure~\ref{fig:uplink} shows that after applying reference image downsampling and uploading the changed tiles in reference images, \name compresses the reference image by over 10000$\times$ and meets the requirement of the limited uplink bandwidth.

\mypara{Computation overhead}
We also measure the runtime that \name and the baselines take to process one imagery.
We use an AMD EPYC 7452 CPU with 16 cores to test the runtime.
As shown in Figure~\ref{fig:computation}, the runtime of \name is the lowest.
Concretely, both \name and the baselines take 0.65 seconds to encode the imagery.
However, Kodan uses an expensive cloud detector that takes 0.39 seconds to run, while \name and SatRoI use the same cheap cloud detector that only takes 0.12 seconds to run.
Further, \name uses \textit{downsampled} reference images to detect changes, allowing the change detection process to run faster than SatRoI's change detection process using full-resolution reference images.

\mypara{More uplink, more improvement}
The performance of \name (in terms of PSNR -- downlink bandwidth trade-off) can be further improved with more uplink capacity, as shown in Figure~\ref{fig:uplink-sensitivity}.
We highlight that \name can further reduce the downlink bandwidth usage by 22 Mbps by increasing the uplink to 4 Mbps.


\mypara{More satellites, more improvement}
\name's \textit{constellation-wide} encoding scheme allows \name to compress satellite imagery with a higher compression ratio when there are more satellites in the constellation.
To verify this, we conduct a simulation based on the Planet thumbnail images (we use thumbnail images to bypass the quota limit of the Planet dataset). We download these images by downloading cloud-free images in Denver from July 1st, 2023 to October 1st, 2023. 
As the pixel values of these thumbnail images is not linearly correlated to the original pixel values sensed by the satellite (due to visual enhancing algorithms like contrast enhancement), we instead calculate the change by normalizing the pixel values to 0-1 and run change detection algorithms on top of these normalized images.
Those changed areas will be downloaded to the ground.
We then calculate the compression ratio based on the average changed areas (\eg 10\% changed areas on average corresponds to 10$\times$ compression)\footnote{
We point out that this is a rough estimation, as in practice image codec's compression efficiency decreases when only encoding a small amount of areas.}.
We plot how the compression ratio (Y-axis) varies with the number of satellites in the constellation (X-axis) in Figure~\ref{fig:more-satellite}.
We highlight that the compression ratio of \name will increase from 3$\times$ to 10$\times$ when the constellation size increases from 1 to 16.

\section{Related Work}
\label{sec:related-work}

\mypara{Single-image compression}
A wide range of prior work has focused on single-image compression by augmenting traditional image codecs like JPEG-2000~\cite{edge-prediction,progressive-jpeg2000,CALIC,linear-prediction,band-reordering,band-reordering-segmentation,band-reordering-prediction,band-prediction} or developing neural-based codec such as autoencoders~\cite{autoencoder-1,autoencoder-2,autoencoder-3,autoencoder-4,autoencoder-5}.
However, this kind of work falls short in leveraging the redundancy \textit{between} images, while satellite imagery remains largely unchanged between two consecutive captures of the same location.

\mypara{Change-based encoding}
A rich set of literature aims to further compress images by detecting changes \textit{between} images.
A line of work builds video-based codecs (\eg H.264~\cite{h264}, H.265~\cite{h265}, VP8~\cite{vp8}, VP9~\cite{vp9} and autoencoders~\cite{dvc,dvc-2,dvc-3}) to leverage such redundancy, with the assumption that two consecutive captures of one location are pixel-wise highly similar.
This is not true for satellite imagery due to varying cloud and illumination conditions. 
Another line of work~\cite{satroi,virtual-background} develops change-based encoding that is robust to varying cloud and illumination conditions.
However, this approach can only update the reference image using single-satellite information.
approach only uses images available for this one location
In contrast, \name allows the satellite to update its reference image using images from \textit{all} the satellites in the same constellation, resulting in a fresher reference image and, thus, better change-based encoding quality.

\mypara{In-orbit computing}
An alternative way to reduce the total downlink capacity is to have a concrete application in mind and drop out images that are irrelevant to this application (\eg Kodan~\cite{kodan} and OEC~\cite{oec}).
However, this approach may drop out images that are crucial for other applications. 
In contrast, \name drops areas that are unchanged, allowing \name to be used by a wider range of applications.



\section{Limitation}
\label{sec:limitation}

While \name improves satellite imagery compression, several concerns remain.

\mypara{Lossy compression}
\name's compression is lossy. While it allows downloading more images, lossy compression may not be applicable to applications that require lossless compression.

\mypara{Control messages}
\name uses the uplink bandwidth that is currently reserved for control messages to upload reference images. 
We believe this is not a serious practical concern as the ground-to-satellite control messages (\eg occasional signaling) do not currently use much of the uplink bandwidth~\cite{limited-uplink}. 

\mypara{Generalization of results}
Our evaluation of \name focuses on a specific set of satellite specs and imagery datasets, but it does not show how effective \name would be if it is used on other or future earth-observation satellites. 
We hope our work will inspire more research to examine \name in other environments.

\mypara{Deployment concerns}
Though \name only changes software, there may be complications in implementing \name on existing satellites as \name requires a software update on the satellite's imagery encoding module onboard the satellite.

Stepping back, we acknowledge that \name does increase the system complexity, especially on the ground stations, including sharing downloaded images across ground stations efficiently. 
However, we believe \name takes an initial step towards constellation-wide sharing of information, a fundamental capability that has broader applications than image compression. 


\section{Conclusion}

\ktedit{
While satellite imagery is useful for a wide range of applications, most of the imagery observed by the satellites is not downloaded to the ground due to limited downlink capacity.
This work presents \name, a new onboard satellite imagery compression system to reduce the downlink bandwidth usage.
\name is the first to make reference-based compression efficient, by enabling constellation-wide sharing of fresh reference images across satellites.
To this end, \name uses 
several techniques to judiciously select and upload reference images under limited uplink capacity.
We show that \name can compress the imagery by 3$\times$ without compromising imagery quality on all bands without using more computation and storage resources, all while staying within existing real-world uplink constraints.
}


\newpage
\bibliographystyle{plain}
\bibliography{reference}

\pagebreak
\appendix

\section{Estimating the storage overhead of \name}
\label{appendix:storage-overhead}

We take the technical specification of DOVEs constellation to estimate this overhead.
We first estimate the space required for storing captured images.
Let the area that the satellite can download during one ground contact be $a$ km$^2$.
The storage space required to store this $a$ km$^2$ imagery is thus $0.87a$ MB, where the coefficient 0.87 is the megabytes required to encode 1km$^2$ area, estimated by the fact that each image captured by Doves constellation is 300 MB, with a resolution of 6600$\times$4400 and a ground sampling distance of 3 $m$.
Thus, the storage space used to store captured imagery is approximately $2 \times 0.87a$ MB, where this 2$\times$ factor is because the ground keeps the captured imagery for two consecutive ground contacts to make sure the downloading is successful~\cite{leoconn}.
\name stores the reference images of all locations that each satellite will download, totaling at most $160a$ km$^2$, since the satellite revisits the same location once every 10-15 days~\cite{sentinel-2-revisit} and the maximum amount of ground contact it can have is 240 times, assuming that the satellite can have ground contact during every 90-minute orbit.
Since \name's downsampling technique compresses the reference images by $2601\times$. As a result, the total storage space for reference images is at most $0.08a MB$, 9\% of the space for storing captured imagery.

\end{document}